\documentclass[aps,prb,twocolumn,superscriptaddress,floatfix]{revtex4-2}
\usepackage[utf8]{inputenc}
\usepackage[english]{babel}
\usepackage{amsmath}
\usepackage{feynmf}
\usepackage{xcolor}
\usepackage{soul}
\usepackage{graphicx}
\usepackage{amsfonts}
\usepackage{blindtext}
\usepackage{appendix}
\usepackage[colorlinks = true,
            linkcolor = blue,
            urlcolor  = blue,
            citecolor = blue,
            anchorcolor = blue]{hyperref}
\usepackage{siunitx}
\usepackage{xcolor}
\usepackage{booktabs}
\usepackage{soul}
\usepackage[normalem]{ulem}

\begin{abstract}
Recent experiments on cuprates have shown the possibility of opening a gap above the superconducting critical temperature, in the so-called phase-fluctuating state, by enhancing the phase coherence of preformed Cooper pairs. Quench-drive spectroscopy, an implementation of 2D coherent spectroscopy, has emerged as a powerful tool for investigating out-of-equilibrium superconductors and their collective modes. 
In this work, we enrich the quench-drive scheme by developing a systematic generalization to study the nonlinear response of $d$-wave incoherent Cooper pairs in a symmetry resolved manner. In particular, we do not only show that it is possible to obtain a third harmonic signal from fully incoherent pairs with an equilibrium vanishing order parameter, but we also characterize the full flourishing 2D spectrum of the generated nonlinear response. The results provide a deeper theoretical insight on recent experimental results, opening the door to a new symmetry-driven design of future experiments on unconventional and enhanced superconductors.

\end{abstract}

\begin{document}

\title{Theory of symmetry-resolved quench-drive spectroscopy: Nonlinear response of phase-fluctuating superconductors}

\author{Matteo Puviani}
\email{matteo.puviani@mpl.mpg.de}
\affiliation{Max Planck Institute for the Science of Light, 91058 Erlangen, Germany}
\date{\today}
\maketitle

\section{Introduction} 

Since their discovery, high-temperature superconductors have been intensely studied because of their properties and rich phase diagram \cite{Wu1987, Maeda_1988, Keimer2015}. These unconventional superconductors are characterized by a complex order parameter whose value depends on the quasiparticles' crystal momentum: it can assume both positive and negative values with maximum absolute value at the antinodal points of the Brillouin zone, while vanishing at the nodal points \cite{Toda2014rotational}. This character is a result of the $B_{1g}$ symmetry of the $d_{x^2-y^2}$ superconducting pairing, descending from their $D_{4h}$ crystal structure \cite{Devereaux1994Electronic, Markiewicz2005}. However, some features of this class of materials are still under debate, such as the conditions and possibility to induce and experimentally observe collective modes \cite{Matsunaga2014, Cea2016nonlinear, Katsumi2018higgs, chu2020phase, Puviani2021PRL, chu2021fano, glier2023direct, cheng2023evidence}, or the origin of the pseudogap phase \cite{Harris1996, Wang2005field, Rourke2011}.  \\
\indent In particular, various attempts have been made to study and detect the amplitude Higgs mode even in unconventional superconductors, both investigating the nonequilibrium nonlinear behavior of these materials \cite{schwarz2020theory, chu2021fano}, and characterizing the symmetries of their response \cite{Cea2018polarization, Schwarz2020}. Recent advances have shown that the nonlinear behavior of unconventional superconductors when probed by light emerges as the blending of different contributions, depending on electron-hole doping and impurity concentration, among others \cite{Udina2022Faraday, AlasRodrguez2022, katsumi2023revealing}. \\
\indent Beside this, it has been suggested that the pseudogap phase is a precursor of the superconducting state, characterized by finite pairing strength and pre-formed Cooper pairs with phase incoherence \cite{Harris1996, Wang2005field, Giusti2019signatures}. Even if this picture is controversial and has been disproved to some extent, in cuprates, in a region of the phase diagram above the superconducting critical temperature, the superconducting phase is incoherent \cite{Emery1995, Corson1999, Xu2000}. \\
\begin{figure*}[ht!]
\centering
\includegraphics[width=17cm]{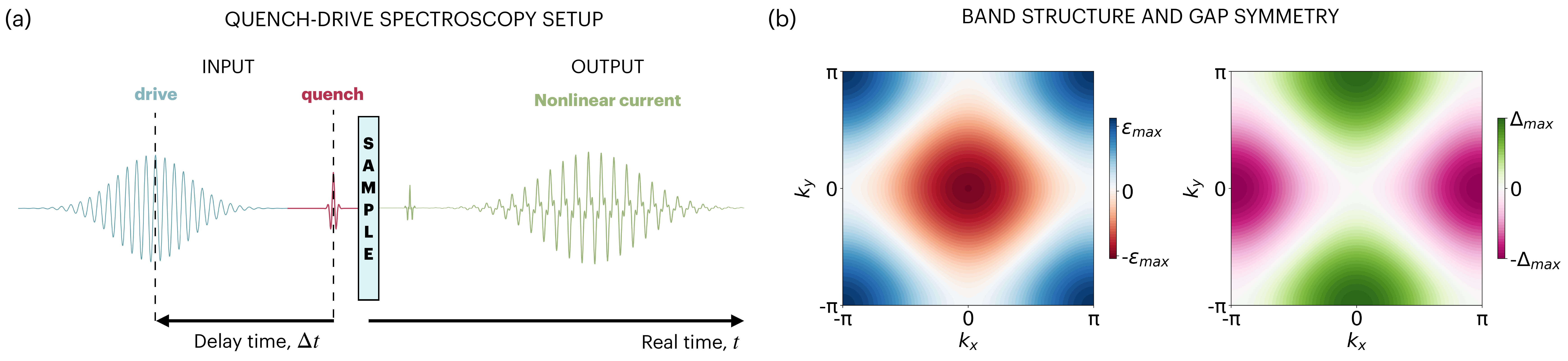}
\caption{Quench-drive spectroscopy of unconventional superconductors. (a) The 2D quench-drive spectroscopy is performed with a short (quench) pulse, followed by a long gaussian-shaped (drive) pulse at a delayed time $\Delta t$. The output signal is analyzed as a function of the real time $t$. (b) Band structure (on the left) and $d_{x^2-y^2}$ gap symmetry (on the right) of the unconventional superconductors studied in this work. The nodal points are identified with $\Delta_\mathbf{k} = 0$, for $\mathbf{k} = (k_x, k_y = \pm k_x)$, while the antinodal ones are at $\mathbf{k} = (0, \pm \pi), (\pm \pi, 0)$.} \label{scheme}
\end{figure*}
When an electromagnetic field interacts with a phase-fluctuating superconductor, for some values of intensity and frequency of the incident radiation, it is possible to induce phase coherence among pre-existing phase-incoherent Cooper pairs: this process is responsible for the transient enhancement of the order parameter, or the appearance of a finite superconducting gap in the case of complete phase-incoherence \cite{Giusti2019signatures, Giusti2021anisotropic}. If this transition from an incoherent to a partially-coherent phase is fast enough, such as when induced by a short-time quench pulse, then oscillations of the order parameter (quasiparticles' and amplitude mode's excitations) can be produced as well, similarly to what happens in light-induced superconductors \cite{Buzzi2021Higgs}. \\
\indent Moreover, in the last years it has been theoretically shown and experimentally observed the generation of odd higher harmonics from driven superconductors: this result originates from the nonlinear behavior of the optical kernel in the superconducting state \cite{Cea2016nonlinear, schwarz2020theory}.
2D coherent spectroscopy (2DCS) on superconductors \cite{Udina2019theory} has developed as a systematic generalization of pump-probe \cite{Giorgianni2019} in the context of the broader concept of high-dimensional spectroscopy \cite{Cundiff2013,Woerner2013,PhysRevLett.118.207204,PhysRevLett.122.257401,Mahmood2021}. In addition, quench-drive spectroscopy (Fig.~\ref{scheme}(a)) has been proposed by Puviani et al. \cite{Puviani2022, Puviani2023} as a specific scenario of THz 2DCS to study superconductors, combining a few-cycle short-time quench pulse and a long-time multi-cycle driving field. This allows to obtain a complex 2D nonlinear response embedding many nonlinear contributions, providing useful information on the optical kernel of the superconductor. In fact, this technique can be used on superconductors to study high-harmonic generation, and to address quasiparticles' excitations as well as collective states, as shown by a recent experimental realization \cite{kim2023tracing}. Since then, these two dimensional spectroscopies have widely developed \cite{Mootz2022visualization, Luo2022, mootz2023twodimensional, mootz2024multidimensional}, proving to be suitable for extracting novel information and details on the superconducting order parameter and its collective modes \cite{katsumi2023revealing, salvador2024principles}. \\
\indent In this work we combine the quench-drive spectroscopy technique, which allows to investigate the non-equilibrium behavior of materials, with the symmetry selection allowed by pulses' polarization typically used in other spectroscopic techniques, such as Raman \cite{Devereaux1994Electronic, Devereaux1995, Devereaux2007inelastic} or birefringence \cite{Giusti2021anisotropic} spectroscopy. Here, we study the nonlinear response to quench-drive pulses of fully phase-incoherent Cooper pairs with $d_{x^2-y^2}$-wave pairing symmetry (Fig.~\ref{scheme}(b)-(c)), as in unconventional superconductors. In particular, we systematically investigate the order parameter's dynamics and the high-harmonic generation process as a function of the real time and the quench-drive delay time, as well as their Fourier spectra. 
The main results of our work can be summarized with the following points:
\begin{enumerate}
    \item presence of non-linear response even with vanishing order parameter at equilibrium;
    \item the induced gap oscillations have predominant $B_{1g}$ and $B_{2g}$ symmetry according to the spectroscopy scheme;
    \item the non-linear current response has $B_{1g}$ or $B_{2g}$ symmetry according whether it originates from the driving or the quench pulse, respectively;
    \item the third harmonic generation originates from predominant $B_{1g}$ symmetry excitation.
\end{enumerate} 
We also want to highlight that all these results are directly experimentally accessible, and the first one
suggests that our approach can be used for testing the hypothesis of incoherent pairs above Tc. In addition, while the first result does not require necessarily a symmetry-resolved quench-drive spectroscopy scheme to be obtained (even if only partial information would otherwise be caught), the other results are only achievable with the use of the symmetry-resolved technique introduced in this work. \\
\indent The paper is organized as follows: in Section \ref{theory} we provide a brief theoretical overview of nonlinear current generation by incoherent Cooper pairs with $d_{x^2-y^2}$ symmetry. In Section \ref{SymmetryResolved} we describe the foundations of symmetry-resolved quench-drive spectroscopy. In Section \ref{results} we show and analyze the numerical results, studying the symmetry-resolved nonlinear response obtained for different configurations of quench and drive pulses. Eventually, in Section \ref{conclusion} we conclude summarizing the work and providing an outlook for possible extensions and future research. In the Appendices \ref{AppendixPseudo} and \ref{AppendixQD} we propose the full theoretical calculations of the pseudospin model and the quench-drive nonlinear response generation, while in Appendix \ref{MoreResults} we provide more results, obtained with a different choice of the quench and drive frequencies.

\section{Nonlinear response of incoherent Cooper pairs} \label{theory}
In this section we theoretically investigate the nonlinear current generated by a material in a state with phase-fluctuating superconductivity subject to quench and drive pulses. The result is obtained by solving the Bloch equations derived from the pseudospin model of the BCS Hamiltonian described in Appendix \ref{AppendixPseudo} and \ref{AppendixQD}. \\
In cuprates, recent experimental results have shown the presence of superconducting fluctuations even above the superconducting critical temperature \cite{Kondo2015}. This behavior has been explained by postulating the presence of incoherent Cooper pairs: in this picture, while the pairing persists even above the critical superconducting temperature, the Cooper pairs lose their phase coherence \cite{Madan2014}. This has been experimentally supported by further photoemission \cite{Reber2013}, magnetization \cite{Yu2019universal} and transport measurements \cite{Pelc2018, Popevi2018}, which suggest the presence of local correlations and superconducting pairing above the critical temperature \cite{Giusti2019signatures}.
Therefore, in order to model the state with phase-fluctuating superconductivity characterized by the presence of pre-formed incoherent pairs, we consider a new artificial equilibrium superconducting state obtained by adding a random momentum-dependent phase $\phi_\mathbf{k}$ to the original Cooper pairs' state, as in Ref. \cite{Giusti2021anisotropic}. As a result, the strength of the pairing potential remains unchanged, as well as the number of total Cooper pairs, while the superconducting order parameter decreases due to the reduced coherence. 
According to the maximum angle $\phi_{max}$ which defines the range of the random phase $\phi_\mathbf{k}$, with $\phi_\mathbf{k} \in [- \phi_{max}, + \phi_{max}]$, we are able to describe different conditions of the material, from the pure superconducting phase for $\phi_{max} = 0$, to the complete loss of coherence for $\phi_{max} = \pi$.\\
We define the gap of the pure superconducting state $\Delta^{(0)}_\mathbf{k} = \Delta^{(0)}_0 f_\mathbf{k}$, and the superconducting order parameter in the presence of incoherent pairs as $\tilde{\Delta}^{(\phi)}_\mathbf{k} = \tilde{\Delta}^{(\phi)} f_\mathbf{k}$, such that  
\begin{align} \label{PGgap}
\tilde{\Delta}^{(\phi)} = V \sum_{\mathbf{k}'} f_{\mathbf{k}'}^2 \dfrac{\Delta_0^{(0)}}{2 E_{\mathbf{k}'}^{(0)}} e^{i \phi_{\mathbf{k}'}} \,,
\end{align}
where $V$ is the same pairing strength of the original state, and $E_{\mathbf{k}}^{(0)} = \sqrt{\epsilon_{\mathbf{k}}^2 + (\Delta^{(0)}_\mathbf{k})^2}$. We notice that the order parameter is calculated with the sum of the coherent contributions over all the Cooper pairs in momentum space. In  phase-fluctuating superconductors, the global coherence is lost as the Cooper pairs acquire an additional momentum-dependent phase $\phi_k$.
The superconducting gap in the new equilibrium state can be written in the pseudospin formalism as \cite{Anderson1958}
\begin{align}
\tilde{\Delta}^{(\phi)}_\mathbf{k} = V f_\mathbf{k} \sum_{\mathbf{k}'} f_{\mathbf{k}'} \left( \tilde{\sigma}_{\mathbf{k}', x} - i \tilde{\sigma}_{\mathbf{k}', y} \right) \,,
\end{align}
where $f_{\mathbf{k}}$ is the $d_{x^2-y^2}$-wave symmetry of the superconducting pairing. Moreover we have introduced the equilibrium pseudospin components
\begin{eqnarray}
    \left\{
    \begin{array}{ll}
    \begin{split}
    \tilde{\sigma}_{\mathbf{k}, x} &= \sigma_{\mathbf{k}, x} \cos{\phi_\mathbf{k}} = f_\mathbf{k} \dfrac{\Delta_0^{(0)} \cos{\phi_\mathbf{k}}}{2 E_{\mathbf{k}}^{(0)}} \,, \\
    \tilde{\sigma}_{\mathbf{k}, y} &= - \sigma_{\mathbf{k}, x} \sin{\phi_\mathbf{k}} = -f_\mathbf{k} \dfrac{\Delta_0^{(0)} \sin{\phi_\mathbf{k}}}{2 E_{\mathbf{k}}^{(0)}} \,, \\
    \hat{\tilde{\sigma}}_\mathbf{k}^z &= \hat{\sigma}_\mathbf{k}^z \,.
    \end{split}
    \end{array}
    \right. \label{PseudoEq}
\end{eqnarray}
In order to describe the dynamics of the system, we use the Heisenberg's equation of motion 
\begin{align} \label{Heis2}
\partial_t \tilde{\sigma}_\mathbf{k} = \tilde{\mathbf{b}} \times \tilde{\sigma}_\mathbf{k} \,,
\end{align}
with the new pseudomagnetic field defined as
\begin{align}
\tilde{\mathbf{b}} = (- 2 \tilde{\Delta}' f_\mathbf{k}, - 2 \tilde{\Delta}'' f_\mathbf{k},  2 \epsilon_{\mathbf{k}}) \,.
\end{align}
In the presence of an external gauge field represented by the vector potential $\mathbf{A}(t)$ coupling to the electrons, the pseudospin changes in time according to
\begin{align}
\tilde{\mathbf{\sigma}}_\mathbf{k} (t) = \tilde{\mathbf{\sigma}}_\mathbf{k} (0) + \delta \tilde{\mathbf{\sigma}}_\mathbf{k} (t) \,.  \label{Pseudo3}
\end{align}
The vector potential is not restricted here to any particular form, but in the context of quench-drive spectroscopy we will describe is as the sum of the quench and drive pulses' contributions, $\mathbf{A}(t) = \mathbf{A}_{q} (t) + \mathbf{A}_{d} (t) = \mathbf{\overline{A}}_{q} (t-t_q) + \mathbf{\overline{A}}_{d} (t-t_d)$. Here $\mathbf{A}_{q(d)} (t)$ is the vector potential of the quench (drive) pulse only, with amplitude $A_{q(d)}$, respectively, while $\mathbf{\overline{A}}_{q(d)}$ is the quench (drive) pulse shape, shifted at center time $t_{q(d)}$, respectively.
The external electromagnetic field is included in the pseudo-magnetic field by means of the Peierls' minimal substitution $\mathbf{k} \rightarrow \mathbf{k} - e \mathbf{A}(t)$ in the fermionic energy, resulting in
\begin{align}
\tilde{\mathbf{b}}_\mathbf{k} (t) = (- 2\tilde{\Delta}'(t) f_\mathbf{k},  - 2 \tilde{\Delta}''(t) f_\mathbf{k},  \varepsilon_{\mathbf{k} - e \mathbf{A}(t)} + \varepsilon_{\mathbf{k} + e \mathbf{A}(t)}) \,.
\end{align}
Here we considered the limit of small superconducting gap velocity (in comparison to the electron velocity), so that the minimal coupling of the pairing term can be neglected \cite{Ghatak2018}.
The equation of motion in Eq.~(\ref{Heis2}) can be decomposed into a set of differential equations, whose solution provides the time-dependent value of the pseudospin $\tilde{\sigma}_\mathbf{k} (t)$. Once this term is known, we can obtain the value of the time-dependent order parameter $\tilde{\Delta}^{(\phi)} (t) = \tilde{\Delta}^{(\phi)} (0) + \delta \tilde{\Delta}^{(\phi)} (t)$, as well as the generated nonlinear current (see Appendix \ref{AppendixQD}). However, we notice that the complex order parameter can be written as
\begin{align}
\tilde{\Delta}^{(\phi)} = |\tilde{\Delta}^{(\phi)}| \ e^{i \theta} \,,
\end{align}
where $\theta$ is the global phase of the superconducting gap. However, an additional momentum-dependent phase appears in the definition of the order parameter according to Eq.(\ref{PGgap}): as a result the gap equation is not self-consistent anymore and the value of the gap is subject to some time-dependent noise due to the phase incoherence of the preformed pairs. \\
In the full generated current, we can distinguish two non-vanishing contributions: namely, a linear component with the same oscillating behavior of the driving field $\mathbf{A}(t)$
\begin{align}
\mathbf{j}^{(1)}(t) = - e^2 \sum_\mathbf{k} \mathbf{A} (t) \cdot \nabla_{\mathbf{k}}  \mathbf{v}_{\mathbf{k}} \ \left( 2 \hat{\tilde{\sigma}}_\mathbf{k}^z (0) + 1 \right) \,, \label{LinearCurr}
\end{align}
and a non-linear term including all higher orders
\begin{align}
\mathbf{j}^{(NL)} (t) = e \sum_\mathbf{k}  \mathbf{v}_{\mathbf{k} - e \mathbf{A}(t)} (2 \hat{\tilde{\sigma}}^z_\mathbf{k} (t) - 2 \hat{\tilde{\sigma}}^z_\mathbf{k} (0))  \,. \label{NLcurrent2}
\end{align}
Since the third pseudospin component in equilibrium is independent on the phase coherence (Eq.~(\ref{PseudoEq})), the linear current in Eq.~(\ref{LinearCurr}) is always nonzero, even for fully incoherent Cooper pairs and vanishing gap. \\
More details on the solution of the equation of motion and the derivation of the generated current for the quench-drive setup are provided in Appendices \ref{AppendixPseudo} and \ref{AppendixQD}.
\begin{figure}[tb!]
\centering
\includegraphics[width=8cm]{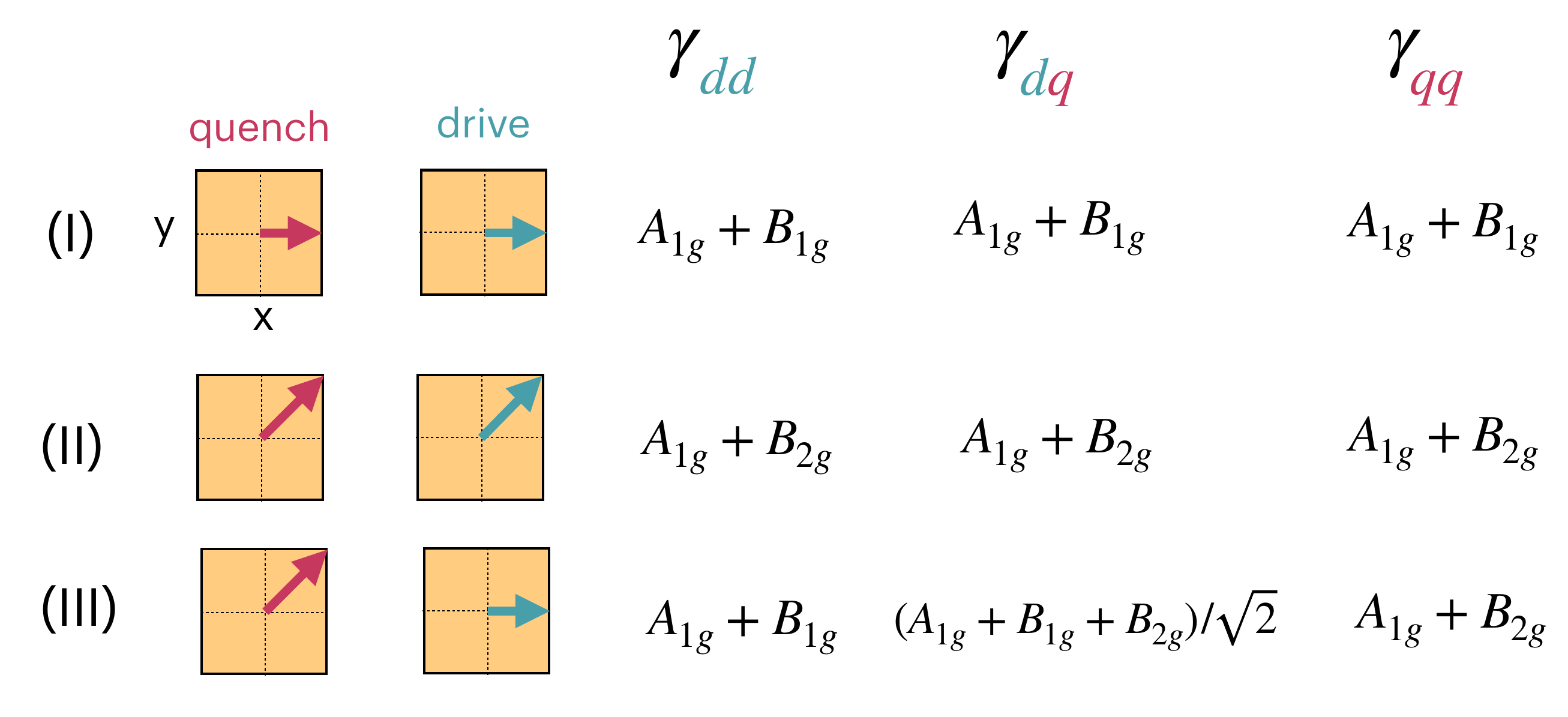}
\caption{Table of symmetries of Raman factors, $\gamma_{rs}$, with $r,s \in \lbrace q, p \rbrace$, where the labels $q$ and $p$ represent the quench and drive pulses' directions, respectively.} \label{table1}
\end{figure}

\begin{figure*}[ht!]
\centering
\includegraphics[width=17cm]{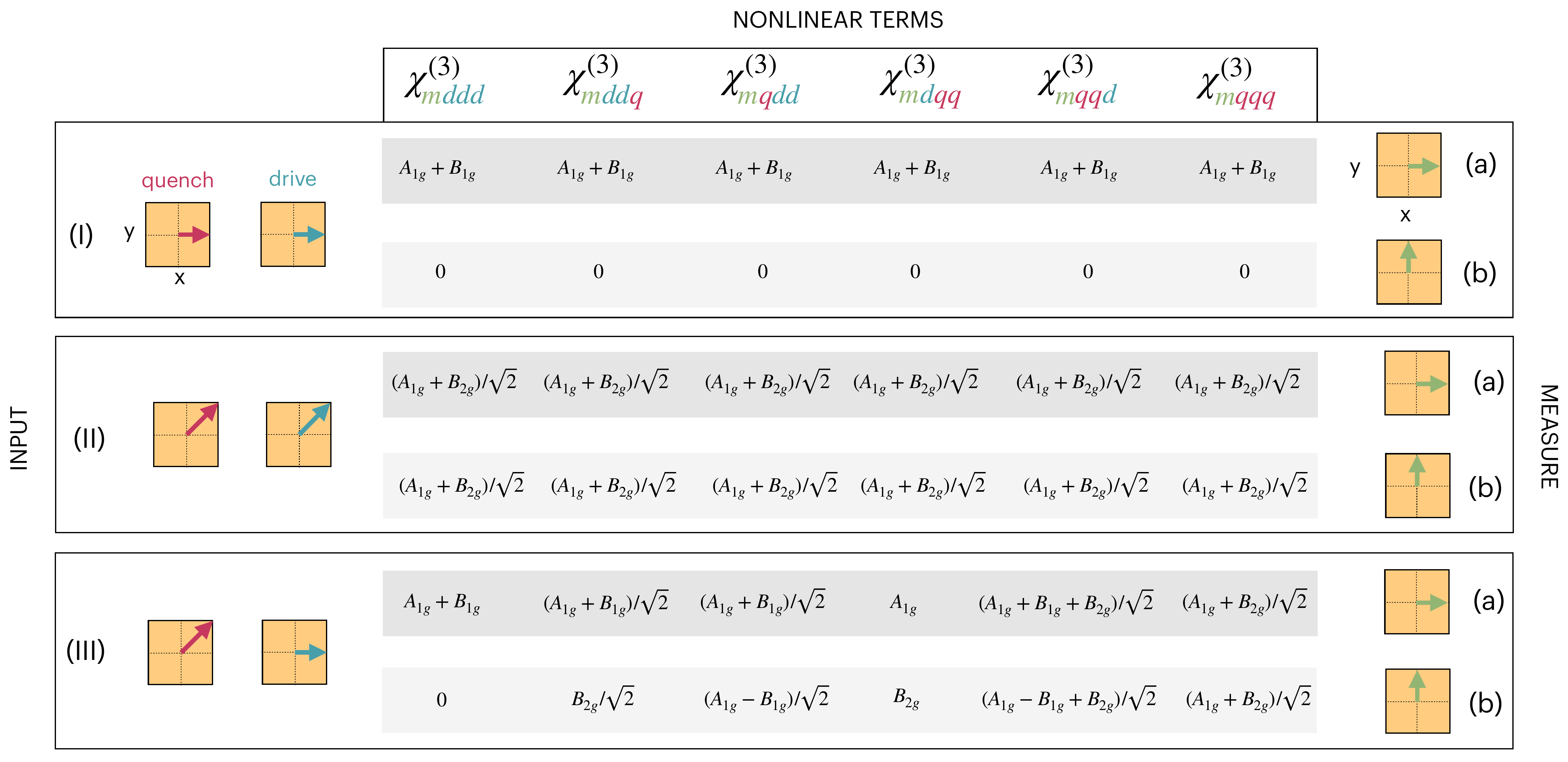}
\caption{Table of symmetry-resolved nonlinear spectra contributions. The table shows the six symmetry-resolved components of the third-order nonlinear susceptibility $\chi^{(3)}_{ijkl} \propto \chi^{(3)}_{\rho \rho} \gamma_{ij} \gamma_{kl}$ (with $ \lbrace i, j, k, l \rbrace \in \lbrace m, q, d \rbrace$, which are the measurement, quench and drive axis, respectively) in a 2D quench-drive spectrum on a $D_{4h}$ crystal.  We considered three (I)-(III) given input (quench and drive) directions as well as two (a),(b) measurement axes.} \label{table2}
\end{figure*}

\section{Symmetry-resolved nonlinear 2D spectroscopy} \label{SymmetryResolved}
In this section we propose the theoretical foundations for the symmetry-resolved quench-drive spectroscopy, identifying the main nonlinear components for different configurations and their corresponding symmetry \cite{Matsunaga2017polarization}. \\ 
First, we can conveniently write the frequency spectrum of gap oscillations, as obtained from the solution of the Bloch equations after transforming into Fourier space, with $(\omega, \nu)$ as the conjugate of the time variables $(t, \Delta t)$, using the convolution operation $(\ast)$ defined as $[B \ast C](x) = \int dy \ B(y) C(x-y)$ \cite{Puviani2023}:
\begin{align}
\delta \Delta_{\mathbf{k}} (\omega, \nu) \propto [ A_i \ast A_j ] (\omega, \nu) \ \gamma_{ij} (\mathbf{k}) \,,  \label{GapSym}
\end{align}
with $i,j \in \lbrace x,y \rbrace$, and the vector potential $A_{i,j}$ including both the quench and the driving fields. The Raman-like factor $\gamma_{ij} (\mathbf{k}) = \nabla_\mathbf{k} (\mathbf{v}_\mathbf{k} \cdot \hat{j})$ (where $\hat{j}$ is the unitary vector along the direction of $j$) represents the second-order light-matter coupling and includes the overall symmetry of the gap oscillations (Fig.~\ref{table1}). In this work, we are considering unconventional superconductors characterized by a $D_{4h}$ crystal symmetry, with $d_{x^2-y^2}$ order parameter.
For this point group symmetry the only relevant irreducible representations (\textit{irreps}) are $A_{1g}$, $B_{1g}$ and $B_{2g}$. Therefore, the Raman-like factors can be decomposed into the \textit{irreps} of the $D_{4h}$ point group as follows \cite{Devereaux1995}:
\begin{subequations}
\begin{align}
\gamma_{xx} & = \gamma_{A_{1g}} + \gamma_{B_{1g}} \,, \\
\gamma_{yy} & = \gamma_{A_{1g}} - \gamma_{B_{1g}} \,, \\
\gamma_{xy} & = \gamma_{B_{2g}} \,, \\
\gamma_{x'x'} & = \gamma_{A_{1g}} + \gamma_{B_{2g}} \,, \\
\gamma_{x'y'} & = \gamma_{B_{1g}} \,, \\
\gamma_{y'y'} & = \gamma_{A_{1g}} - \gamma_{B_{2g}} \,, \\
\gamma_{x'x} & = ( \gamma_{A_{1g}} + \gamma_{B_{1g}} + \gamma_{B_{2g}} ) / \sqrt{2} \,, \\
\gamma_{x'y} & = ( \gamma_{A_{1g}} - \gamma_{B_{1g}} + \gamma_{B_{2g}} ) / \sqrt{2} \,,
\end{align}
\end{subequations}
with $\hat{x}' = (\hat{x} + \hat{y}) / \sqrt{2}$ and $\hat{y}' = (\hat{x} - \hat{y}) / \sqrt{2}$, corresponding to an angle with respect to the $\hat{x}$ axis of $\pi / 4 $ and $- \pi / 4$, respectively. The general rule given the angles $\alpha$ and $\theta$ with respect to the $\hat{x}$ axis reads \cite{Giusti2021anisotropic}:
\begin{align}
\gamma_{\alpha \theta} (\mathbf{k}) & = \gamma_{A_{1g}} \cos{(\alpha - \theta)} + \gamma_{B_{1g}} \cos{(\alpha + \theta)} \nonumber \\
& + \gamma_{B_{2g}} \sin{(\alpha + \theta)} \,.
\end{align}
Similarly, the third-order nonlinear current, which represents the lowest-order non-vanishing nonlinear contribution, can be written as
\begin{align}
j_i^{(3)} (t, \Delta t) \propto A_j (t, \Delta t) \sum_\mathbf{k} \gamma_{ij} (\mathbf{k}) \ \delta \tilde{\sigma}^z_\mathbf{k} (t, \Delta t) \,, \label{jSym1}
\end{align}
with $i,j,k,l \in \lbrace x,y \rbrace$. We notice that in this expression the order parameter's oscillations of Eq.(\ref{GapSym}) are embedded into the time-dependent pseudospin component $\delta \tilde{\sigma}^z_\mathbf{k} (t, \Delta t)$.
It is convenient to consider its spectrum in Fourier space, where $(\omega, \nu)$ are the conjugate of the variables $(t, \Delta t)$, as 
\begin{align}
j_i^{(3)} (\omega, \nu)  & \propto \sum_{j,k, l} \int d\omega_1 d \omega_2 d \omega_3 \ \chi_{ijkl}^{(3)} (\omega - \omega_1) \ A_j (\omega_1)   \nonumber \\
& \cdot A_k (\omega_2) \  A_l (\omega_3) \ \delta (\omega - \omega_1 - \omega_2 - \omega_3) \,, 
\label{jSym2}
\end{align}
where the delta function over the frequencies enforces energy conservation.
Here we also introduced the third-order nonlinear susceptibility $\chi_{ijkl}^{(3)} = \gamma_{ij} \ \gamma_{kl} \ \chi_{\rho \rho}^{(3)}$, where $\chi_{\rho \rho}^{(3)}$ is the third-order density-density response function.
\begin{figure*}[tb!]
\centering
\includegraphics[width=17cm]{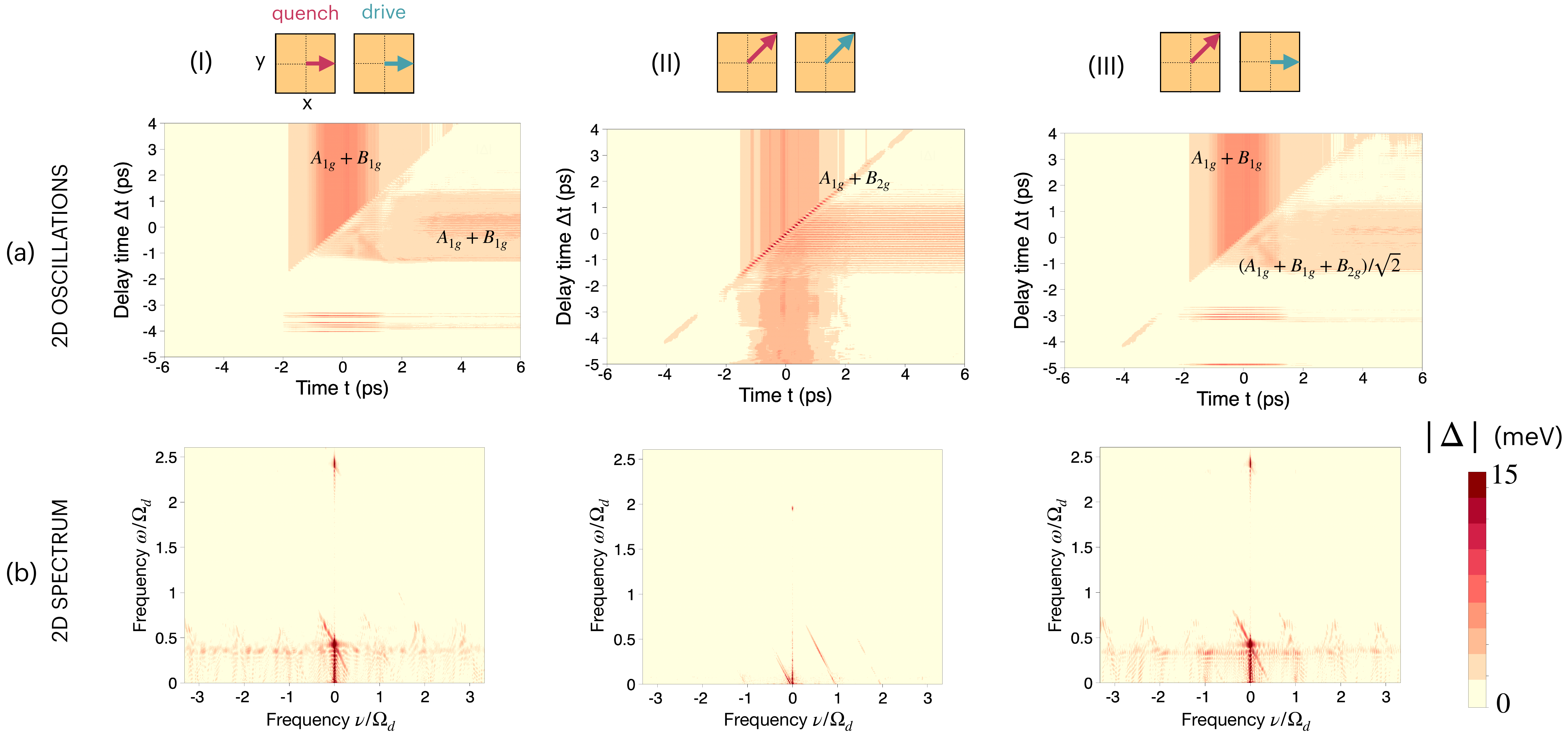}
\caption{Gap oscillations and frequency spectra. (a) 2D oscillations in $(t, \Delta t)$of the absolute value of the superconducting gap, $|\Delta|$, for the three scheme configurations (I)-(III) described in the main text and illustrated by the plots of quench and drive pulses. The main symmetry contributions are written for the strongest signals, according to the table in Fig.\ref{table1}. (b) Absolute value of the 2D Fourier transform of the full complex gap, $| \mathcal{F} \lbrace \Delta  (t, \Delta t) \rbrace| = |\Delta  (\omega, \nu)|$.} \label{Result1}
\end{figure*}
Here we omitted the sum over $\mathbf{k}$ and the frequency dependencies for convenience of notation. \\
As an example, we can derive the symmetry of the $xyx'y$ response, which enters the nonlinear current along the $x$ direction with interaction of pulses along $\hat{x}'$ and $\hat{y}$ , as follows:
\begin{align}
\chi_{xyx'y}^{(3)} & = \chi_{\rho \rho}^{(3)} \ \gamma_{xy} \ \gamma_{x'y} \nonumber \\
& = \chi_{\rho \rho}^{(3)} \ \gamma_{B_{2g}} \ ( \gamma_{A_{1g}} - \gamma_{B_{1g}} + \gamma_{B_{2g}} ) / \sqrt{2} \nonumber \\
& = \chi^{(3)}_{B_{2g}}/ \sqrt{2} \,, \label{exampleChi}
\end{align}
with $\chi_{B_{2g}}^{(3)} = \chi_{\rho \rho}^{(3)} \ \gamma_{B_{2g}} \ \gamma_{B_{2g}} / \sqrt{2}$, which is the only non-vanishing term in Eq.(\ref{exampleChi}) after summing over the full Brillouin zone. When analyzing the quench-drive spectra, we can substitute the subscripts $m$, $q$ and $d$ representing the measurement, quench and drive \textit{axis}, respectively, to the Eq.s (\ref{GapSym}), (\ref{jSym1}) and (\ref{jSym2}). \\
\indent As shown by Puviani et al. \cite{Puviani2023}, there are six contributions of the third-order nonlinear susceptibility in a quench-drive spectroscopy setup, which sum up to provide the full nonlinear response, namely (Fig.~\ref{table2}):
\begin{align}
\chi^{(3)} & = \chi^{(3)}_{mddd} + \chi^{(3)}_{mddq} + \chi^{(3)}_{mqdd} \nonumber \\
& + \chi^{(3)}_{mdqq} + \chi^{(3)}_{mqqd}+ \chi^{(3)}_{mqqq} \,.
\end{align}
Each of them can be decomposed into the $D_{4h}$ symmetry \textit{irreps} as shown before. For example, selecting the output along the $x$ axis parallel to the driving field, and with a quench pulse along the $\hat{xy}$ diagonal, i.e. $m = x$, $q = x'$, $d = x$, for the purely driving response we get:
\begin{align}
\chi_{mddd}^{(3)} & = \chi_{\rho \rho}^{(3)} \ \gamma_{md} \ \gamma_{dd} \nonumber \\
& = \chi^{(3)}_{A_{1g}} + \chi^{(3)}_{B_{1g}} \,,
\end{align}
where the first Raman factor represents the measurement-driving vertex, while the second corresponds to the driving-driving one.
Analogously, for the mixed quench-drive response quadratic in the quench amplitude field we have
\begin{align}
\chi_{mdqq}^{(3)} & = \chi_{\rho \rho}^{(3)} \ \gamma_{md} \ \gamma_{qq} \nonumber \\
& = \chi_{A_{1g}}^{(3)} \,,
\end{align}
and
\begin{align}
\chi_{mqqd}^{(3)} & = \chi_{\rho \rho}^{(3)} \ \gamma_{mq} \ \gamma_{qd} \nonumber \\
& = \left( \chi_{A_{1g}}^{(3)} + \chi_{B_{1g}}^{(3)} + \chi_{B_{2g}}^{(3)} \right) / \sqrt{2} \,,
\end{align}
which will appear at $\nu \neq 0$ in the two-dimensional quench-drive Fourier spectrum of the nonlinear response. 
\indent Interestingly, this example shows in practice how the nature of the two-dimensional spectroscopy allows to extract an $A_{1g}$ symmetry response (and similarly the $B_{1g}$ and $B_{2g}$) from only one susceptibility component. 
Moreover, the presence of multiple contributions for different values of the 2D frequency components $(\omega, \nu)$ allows to measure and selectively address all the symmetries response with only one experiment.

\begin{figure*}[tb!]
\centering
\includegraphics[width=18cm]{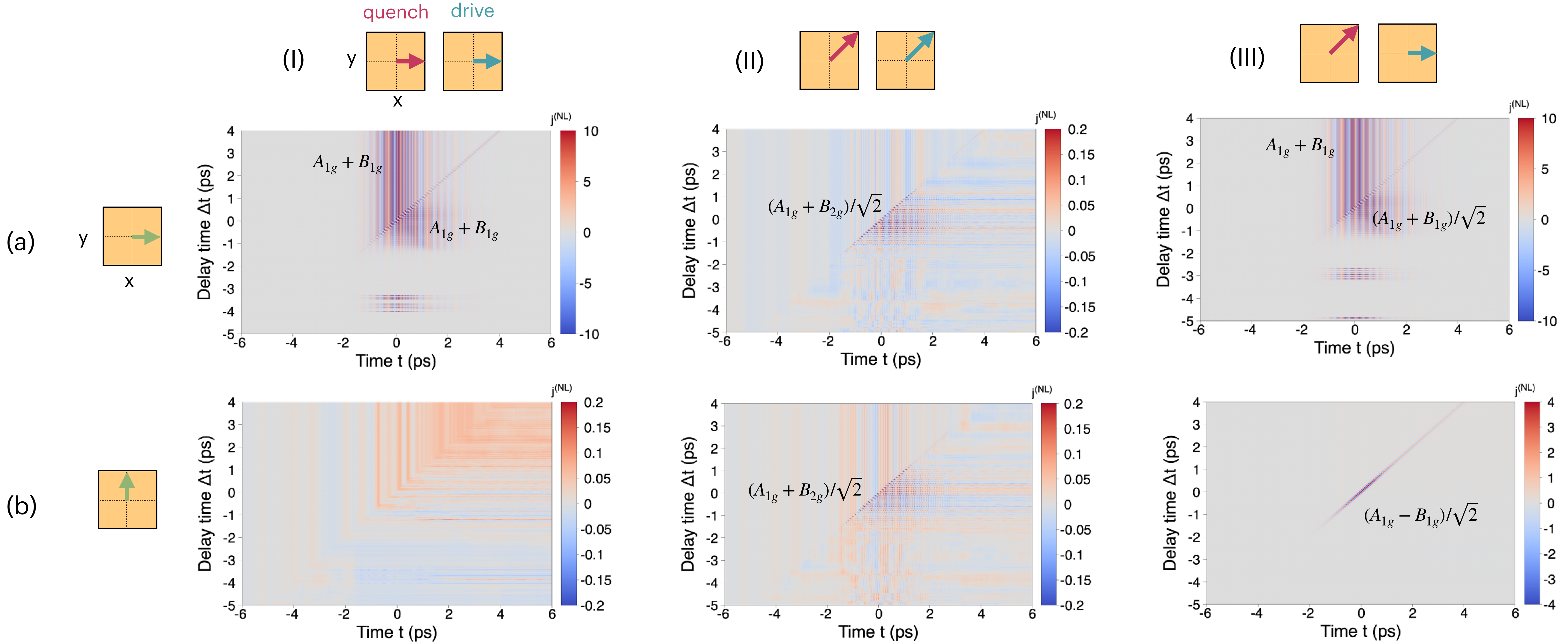}
\caption{2D nonlinear current. Plots of the generated nonlinear current as a function of real time $t$ and the quench-drive delay time $\Delta t$, for three different schemes (I)-(III) described in the main text, and two polarized output measures along (a) $x$ and (b) $y$ axis, respectively. The symmetries written represent the main contribution, according to the table of the nonlinear susceptibilities in Fig.~\ref{table2}. Be aware of the different color scale for each plot.} \label{Result2}
\end{figure*}

\section{Numerical results and discussion} \label{results}
In this section we present the results obtained from the numerical implementation of the expressions and time-dependent Bloch equations show in Sec.~\ref{theory} and \ref{SymmetryResolved}. 
We modeled the electronic band dispersion of the unconventional superconductor as $\epsilon_{\mathbf{k}} = -2 t (\cos{k_x} + \cos{k_y}) - \mu $, where the quasimomentum components are expressed in units of the lattice constant $a$. We used the values of $t = 125$ meV for the nearest-neighbour hopping energy, chemical potential $\mu / t = -0.2$, obtaining an electron occupation $n = 0.9$ as in Ref.~\cite{Giusti2019signatures}.
For the $d_{x^2-y^2}$ order parameter $\Delta_\mathbf{k}^{(0)} = \Delta_{max}^{(0)} (\cos k_x - \cos k_y )/2$ we chose the value $\Delta_{max}^{(0)} = 31$ meV. The calculations were performed with a summation over the full Brillouin zone $\lbrace k_x, k_y \rbrace \in \lbrace - \pi, \pi \rbrace$ with a homogeneous square sampling and a total number of k points $N_\mathbf{k} = 10^6$. For the time-dependent evolution we used a time-step of $\delta t = 3 \cdot 10^{-4}$ ps, and for the quench-drive delay $\delta \Delta t = 2.5 \cdot 10^{-2}$ ps.
For the pulses we used a few-cycle quench and a gaussian-shaped long-duration drive, with vector field amplitudes for the quench and drive pulses $A_q = 0.8$ and $A_d = 0.8$, respectively. Both of the pulses have been described by sinusoidal functions with frequencies $\Omega_q$ and $\Omega_d$, respectively, and gaussian envelopes of shape $e^{-(t-t_{q,d})/(2 \sigma_{q,d}^2)}$ with $2 \sigma_q^2 = 0.01$ ps$^2$ and $2 \sigma_d^2 = 5$ ps$^2$, respectively. Moreover we set the reference time $t=0$ at the center of the gaussian envelope of the driving pulse.
The maximum intensity used for each pulse is provided for the corresponding vector potential in units of $\hbar / (e \ a)$, where $e$ is the electron charge and $a$ the lattice constant. Moreover, we chose the frequency of quench ($\Omega_q$) and drive ($\Omega_d$) to be different but both in the THz spectrum, with values $\Omega_d = 11$ THz and $\Omega_q = 7$ THz. In general, different choices of amplitude and frequencies can be made in order to suppress or enhance specific symmetry contributions.
In this work, we focused on the fully phase-incoherent Cooper pairs, with $\phi_{max} = \pi$, for which $\tilde{\Delta}^{(\phi)} = 0$. \\
In our calculations and analysis we restricted ourselves to only three quench-drive symmetry configurations. These can be addressed in terms of the quench and drive angles defined with respect to the $\hat{x}$ axis $\alpha_q$ and $\alpha_d$, respectively, as follows:
\begin{subequations}
\begin{align}
(I) \quad \alpha_q & = 0 \ \,, \quad \alpha_d = 0 \,, \\
(II) \quad \alpha_q & = \pi/4 \ \,, \quad \alpha_d = \pi/4 \,, \\
(III) \quad \alpha_q & = \pi/4 \ \,, \quad \alpha_d = 0 \,.
\end{align}
\end{subequations}
We studied the behavior of the superconducting gap (Sec. \ref{Gapresults}) and the generated nonlinear current (Sec. \ref{NLresults}) along the $(a) \ x$ and $(b) \ y$ direction, as well as their corresponding 2D spectra. \\
\indent More results, obtained with drive and quench frequencies $\Omega_d = 3.66$ THz and $\Omega_q = 7.48$ THz, respectively, are provided in Appendix \ref{MoreResults}: in this case the quench pulse is nearly resonant with the bare superconducting gap, while the drive is at a much lower energy.

\begin{figure*}[ht!]
\centering
\includegraphics[width=18cm]{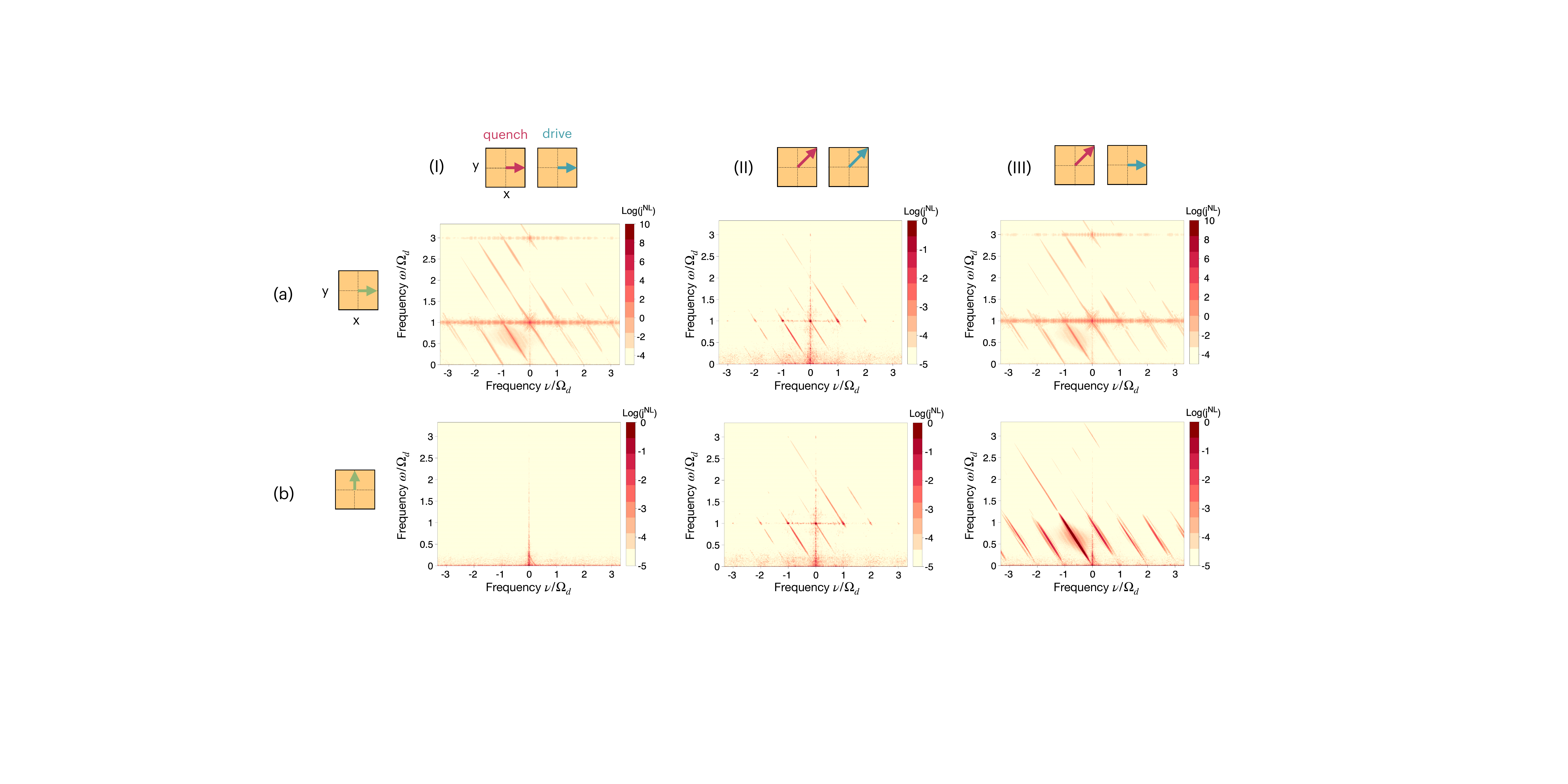}
\caption{2D nonlinear current spectra. Plots of the Fourier transform of the nonlinear current in Fig.\ref{Result2}, for three different schemes (I)-(III) described in the main text, and two polarized output measures along (a) $x$ and (b) $y$ axis, respectively. Be aware of the different Log color scale for each plot. } \label{Result3}
\end{figure*}

\subsection{Emergent superconducting gap and oscillations} \label{Gapresults}
At first, we calculated the behavior of the order parameter, i.e. the superconducting gap, within the quench-drive spectroscopy setup.
In Fig.~\ref{Result1} (I)-(III) (a) we show the 2D time-dependent behavior of the absolute value of the gap. In all our simulations we have set the reference time $t=0$ at the center of the gaussian envelope of the driving pulse. Since the initial system is formed by incoherent pairs, the initial superconducting gap is zero. However, when the quench and drive pulses perturb the incoherent state, they are able to induce coherence in the Cooper pairs, giving rise to a finite gap value, in accordance with Ref.~\cite{Giusti2021anisotropic}. However, thanks to the quench-drive spectroscopic technique, exploiting the symmetry resolution for different quench and drive directions, we can analyze more in depth the gap behavior and the symmetry of its oscillations. Indeed, in Ref.~\cite{Giusti2019signatures} it was shown that a quench pulse along the $x$ axis, i.e. with $\alpha_q = 0$, tends to reduce the superconducting gap decreasing coherence, while with $\alpha_q = \pi/4$ it is increased. Here, we go beyond that scheme observing that, with the given frequencies of the pulses, a long driving pulse with $\alpha_d = 0$ can also induce coherence in a fully incoherent setup, while a quench along the same direction keeps suppressing it (plots (I)(a) and (III)(a) of Fig.~\ref{Result1}). On the other hand, a long driving pulse with $\alpha_d = \pi/4$ can also increase the gap coherence, but with less efficiency (plot (II)(a)). In order to understand this, we can use the symmetry table in Fig.~\ref{table1}: in fact, for both the (I)(a) and (III)(a) conditions, the pairs are excited mainly in the $B_{1g}$ symmetry channel. On the contrary, in the (II)(a) scheme the gap is excited with a predominant $B_{2g}$ symmetry. As a consequence, the $B_{1g}$ symmetry enhances the gap if used in a driving, while it tends to suppress it if imposed by a short quench. \\
Additional information can be extracted from the analysis of the 2D Fourier spectra of the complex gap, as shown in Fig.~\ref{Result1}(b). On the one hand there is a $2 \Omega_d$ oscillation for the (II) scheme, which results from the $B_{2g}$ excitation, while no $2 \Delta$ peak (originating from quasiparticles' and amplitude mode excitations) appears here. On the other hand, in the schemes (I) and (III), where the $B_{1g}$ symmetry is mainly excited as the relevant one, we notice dominant frequency components at $\omega \approx 0.5 \ \Omega_d$ and $\omega \approx 2.5 \ \Omega_d$. The reason is that within this symmetry the dominant excitation of the superconducting gap is provided by the quasiparticles' excitation and amplitude mode, which have an intrinsic frequency of $\omega = 2 \tilde{\Delta}^{(\phi)}$ and $\omega \approx 0.4 \ \tilde{\Delta}^{(\phi)}$, as predicted in Ref.~\cite{Schwarz2020}.

\subsection{Nonlinear current generation} \label{NLresults}
Since the order parameter is not easily accessible in a direct way in experiments, we analyze here the generated nonlinear current by the material: this is because the linear current contains a strong response from the incoherent pairs, while the interesting information is contained in the purely nonlinear part. In Fig.\ref{Result2} we show the 2D current and the corresponding spectra for the (I)-(III), (a)-(b) configurations, indicating the main symmetry contributions to each term, obtained from Fig.\ref{table2}. \\
We first notice that the current measured along the $x$ direction (sub-plots (a)) follows the behavior of the gap in Fig.\ref{Result1}, even though the intensity peak for the $B_{1g}$ symmetries (I),(III) is one order of magnitude larger than the one with $B_{2g}$ (II). This can be partly ascribed to the pulse duration and the frequency difference between the quench and the drive, even though the corresponding gap intensities are in the opposite order. \\
The calculations performed selecting the polarization along the $y$ axis are particularly interesting: in fact the response of configuration (I)(b) vanishes (in accordance to the symmetry-resolved susceptibility in Fig.~\ref{table2}), and the response in (II)(b) is surprisingly lower than the one in (III)(b), even though the gap for $t = \Delta t$ in the latter case is smaller than in the former. We can also notice that the response in (II)(b) occurs only when quench and drive overlap and extends along the $t$ axis, while the current in (III)(b) is visible only along the diagonal $t = \Delta t$, starting when the driving overlaps with the quench. This means that in the former case the $\chi_{mdqq}$, $\chi_{mddq}$ and $\chi_{mqqd}$ are the most relevant contributions, while in the latter $\chi_{mqdd}$ is, with $B_{2g}$ and $B_{1g}$ dominant symmetry, respectively. 
Overall, the $B_{2g}$ symmetry is responsible for the gap enhancement from a short pulse, while the $B_{1g}$ symmetry dominates when a long driving is applied, as well as in the nonlinear current generation. \\
Additional information can be extracted from the 2D spectra, obtained with the Fourier transform of the time-dependent plots (Fig.\ref{Result3}). In general, the signals at $\nu = 0$ are independent of the quench pulse, while all the diagonal lines originate from at least a quench pulse component. The horizontal lines with $\omega = const.$, which appear in I(a) and III(a) in correspondence of the first harmonic signal, are also independent on the $\omega$ frequency and are generated by $\chi_{mddd}^{(3)}$. \\
We first notice that, while in the (II)(b) scheme the most prominent features are peaks at $(\omega = \Omega_d, \nu = n \Omega_d)$ followed by diagonal spectral lines, in (III)(b) the diagonal features peaked below $\omega = \Omega_d$ are more visible, at $\omega \approx \Omega_q, \nu \approx \Omega_q$. In particular, the diagonal signal starting from the origin and with $\omega = -\nu$ is the sum of the contributions of nonlinear susceptibilities $\chi^{(3)}_{mqdd}+\chi^{(3)}_{mddq}+\chi^{(3)}_{mqqq}$. \\
The third harmonic generated by the driving pulse, appearing along the vertical axis for $\nu =0$, is generated by the third-order nonlinear susceptibility $\chi_{mddd}^{(3)}$, and appears in Fig.~\ref{Result3} I(a) and III(a). Its importance is twofold: firstly, this generally proves that it is possible to generate a third harmonic response even in fully incoherent Cooper pairs exhibiting an initially null gap, when properly quenched and driven. This feature has been experimentally shown in cuprate superconductors above their critical temperature, where a phase-fluctuating phase with vanishing gap is expected \cite{chu2020phase}. Secondly, the third harmonic is generated only when the $B_{1g}$ symmetry is explicitly excited (see also Fig.~\ref{table2}). 
However, we can also notice that in configuration (III)(b) there is a non-vanishing third-harmonic component at $\nu = 0$, originating from a diagonal line which accidentally overlaps with $\nu = 0$ due to a higher-order quench-drive mixing, of the kind $\chi^{(5)}_{mq,qddd}$. \\
\indent In addition to this analysis for fully-incoherent Cooper pairs, we extended the same approach to partially-incoherent superconductors (see Appendix \ref{MoreResults} for the results and a detailed analysis). We have shown that the main features of the symmetry-resolved nonlinear quench-drive spectroscopy are still in place, since they are determined by the symmetry of the underlying Cooper pairing. This would allow to point whether there are pre-formed pairs even with a vanishing superconducting order parameter. Moreover, it would be possible to discriminate between a fully-incoherent and a partially-incoherent superconductor by analysing the intensity of the third harmonic signal and especially at its time- and frequency-dependent modulations for given symmetries.

\section{Conclusion and outlook} \label{conclusion}
In this work we have calculated the nonlinear response of a phase-fluctuating superconductor with $d_{x^2-y^2}$ pairing symmetry without phase coherence, characterized by vanishing superconducting gap in equilibrium. We have adopted the recently proposed quench-drive spectroscopy scheme \cite{Puviani2022, Puviani2023}, with THz pulses, inducing a finite superconducting gap and analyzing the generated nonlinear current response. In particular, we have developed a symmetry-resolved analysis, which allows to selectively address symmetry components according to the quench and drive pulses and the measurement axis chosen. \\
We have found (i) a non-linear response even with zero equilibrium order parameter and (ii) induced gap oscillations with predominant $B_{1g}$ and $B_{2g}$ symmetry, according to the spectroscopy scheme. Moreover, (iii) the non-linear current response has $B_{1g}$ or $B_{2g}$ symmetry according whether it originates from the driving or the quench pulse, respectively, while (iv) the third harmonic generation originates from predominant $B_{1g}$ symmetry excitation. We stress that the results (ii)-(iv) explicitly require the symmetry-based technique introduced in this paper to be obtained, giving one demonstration of the power and relevance of symmetry-resolved quench-drive spectroscopy.
\\
We also highlight that our theoretical approach can be applied to any superconductor and superconducting-related effect, since it is based on the redundancy symmetry breaking to generate a nonlinear response. Moreover, the symmetry-resolved analysis is extremely powerful as it allows to identify the underlying symmetry of the order parameter, the Cooper pairing and any physical mechanism giving rise to a photon-induced response. Among these, we mention the possibility to address different collective excitations \cite{Udina2019theory, Gabriele2021}, and even helping in shining light on the superconducting diode effect \cite{Nadeem2023}.

\acknowledgements 
Fruitful discussions with P. M. Bonetti, R. Haenel, S. Kaiser, D. Manske and D. Vilardi are thankfully acknowledged.

\appendix

\section{Pseudospin model for a superconductor} \label{AppendixPseudo}
In this appendix we provide a detailed description of the usage of the pseudospin model to solve the equation of motion of a superconductor when perturbed by an external field. 
In order to describe the superconducting phase of a material, we adopt the BCS model expressed by the mean field Hamiltonian
\begin{align} \label{BCSHam}
\hat{H}_{BCS} = \sum_{\mathbf{k}, \sigma} \epsilon_{\mathbf{k}} \hat{c}_{\mathbf{k}, \sigma}^\dagger \hat{c}_{\mathbf{k}, \sigma} - \sum_{\mathbf{k}} \left( \Delta_{\mathbf{k}}\hat{c}_{\mathbf{k}, \uparrow}^\dagger \hat{c}_{-\mathbf{k}, \downarrow}^\dagger + h.c. \right) \,,
\end{align}
where $\epsilon_{\mathbf{k}} = \xi_\mathbf{k} - \mu$, $\xi_\mathbf{k}$ is the electronic band dispersion, $\mu$ the chemical potential and $\Delta_{\mathbf{k}}$ the momentum-dependent superconducting order parameter. This latter is described by a complex number which satisfies the gap equation
\begin{align} \label{gapEq}
\Delta_\mathbf{k} = \sum_{\mathbf{k}'} V_{\mathbf{k},\mathbf{k}'} \langle \hat{c}_{-\mathbf{k}', \downarrow} \hat{c}_{\mathbf{k}', \uparrow} \rangle \,,
\end{align}
$V_{\mathbf{k}, \mathbf{k}'}$ being the (momentum-dependent) pairing interaction. It can be factorized as $V_{\mathbf{k}, \mathbf{k}'} = V f_\mathbf{k} f_{\mathbf{k}'}$, with $f_\mathbf{k} = f^{(d_{x^2-y^2})}_\mathbf{k} = (\cos k_x - \cos k_y)/2$ the $d$-wave form factor of the superconducting order parameter. Therefore, it follows from Eq.~(\ref{gapEq}) that the gap function itself can be factorized as $\Delta_{\mathbf{k}}= \Delta_0 f_\mathbf{k} $. \\
We now rewrite the BCS Hamiltonian using the pseudospin formalism as \cite{Anderson1958, Tsuji2015theory, schwarz2020theory}
\begin{align}
\hat{H}_{BCS} = \sum_\mathbf{k} \mathbf{b}_\mathbf{k} \cdot \hat{\mathbf{\sigma}}_\mathbf{k} \,,
\end{align}
with the pseudospin vector
\begin{align}
\hat{\mathbf{\sigma}}_\mathbf{k} = \frac{1}{2} \hat{\Psi}_\mathbf{k}^\dagger \mathbf{\tau} \hat{\Psi}_\mathbf{k} \,,
\end{align}
which is defined in Nambu-Gor'kov space, with spinor $\hat{\Psi}_\mathbf{k}^\dagger = (\hat{c}_{\mathbf{k}, \uparrow}^\dagger \quad \hat{c}_{- \mathbf{k}, \downarrow})$ and the Pauli matrices $\mathbf{\tau} = (\tau_1, \tau_2, \tau_3)$. The pseudo-magnetic field is defined by the vector
\begin{align}
\mathbf{b}_\mathbf{k} = (- 2 \Delta' f_\mathbf{k}, - 2 \Delta'' f_\mathbf{k}, 2 \epsilon_{\mathbf{k}}) \,,
\end{align}
where $\epsilon_{\mathbf{k}} = \xi_\mathbf{k} - \mu$, $\xi_\mathbf{k}$ being the fermionic band dispersion, $\mu$ the chemical potential. 

In the presence of an external gauge field represented by the vector potential $\mathbf{A}(t)$ coupling to the electrons, the pseudospin changes in time according to
\begin{align}
\mathbf{\sigma}_\mathbf{k} (t) = \mathbf{\sigma}_\mathbf{k} (0) + \delta \mathbf{\sigma}_\mathbf{k} (t) \,, 
\end{align}
with $\delta \mathbf{\sigma}_\mathbf{k} (t) = (x_\mathbf{k} (t), y_\mathbf{k} (t),  z_\mathbf{k} (t))$. The external electromagnetic field is included in the pseudo-magnetic field by means of the minimal substitution $\mathbf{k} \rightarrow \mathbf{k} - e \mathbf{A}(t)$ in the fermionic energy, resulting in
\begin{align}
\mathbf{b}_\mathbf{k} (t) = (- 2 \Delta'(t) f_\mathbf{k},  - 2 \Delta''(t) f_\mathbf{k},  \varepsilon_{\mathbf{k} - e \mathbf{A}(t)} + \varepsilon_{\mathbf{k} + e \mathbf{A}(t)}) \,.
\end{align}
The Heisenberg equation of motion for the pseudospin can be written in the Bloch form, $\partial_t \mathbf{\sigma}_{k} = \textbf{b}_{\textbf{k}} \times \mathbf{\sigma}_{\textbf{k}}$, providing the set of differential equations
\begin{eqnarray}
    \left\{
    \begin{array}{ll}
    \begin{split}
        \partial_t x(t) &= - (\varepsilon_{\mathbf{k} - e \mathbf{A}} + \varepsilon_{\mathbf{k} + e \mathbf{A}}) y(t) - \dfrac{f_\mathbf{k}}{E_{\mathbf{k}}} \epsilon_{\mathbf{k}} \delta \Delta'' (t)  \\
        &+ 2 \delta \Delta'' (t) f_\mathbf{k} z(t) \,, \\
        \partial_t y(t) &= 2 \varepsilon_{\textbf{k}} x(t) + 2 ( \Delta + \delta \Delta' (t) ) f_\mathbf{k} z(t) \\
        &- \ \delta \Delta' f_\mathbf{k} \dfrac{\epsilon_{\mathbf{k}}}{E_{\mathbf{k}}} + \dfrac{\Delta f_\mathbf{k}}{2 E_{\mathbf{k}}} (\varepsilon_{\mathbf{k} - e \mathbf{A}} + \varepsilon_{\mathbf{k} + e \mathbf{A}} - 2 \epsilon_{\mathbf{k}}) \,, \\
        \partial_t z(t) &= -2 \ \Delta f_\mathbf{k} \ y(t) - \dfrac{\Delta f_\mathbf{k}^2}{E_{\mathbf{k}}} \delta \Delta'' (t) - 2 \delta \Delta'' (t) f_\mathbf{k} x(t) \,.
    \end{split}
    \end{array}
    \right. 
    \label{Blocheqs}
\end{eqnarray}
Here, for simplicity of calculations and without loss of generality, we assumed a real order parameter, $\Delta'' (t=0) = 0$, at the initial time $t=0$, so that $y(0) = 0$. 

\begin{figure*}[tb!]
\centering
\includegraphics[width=18cm]{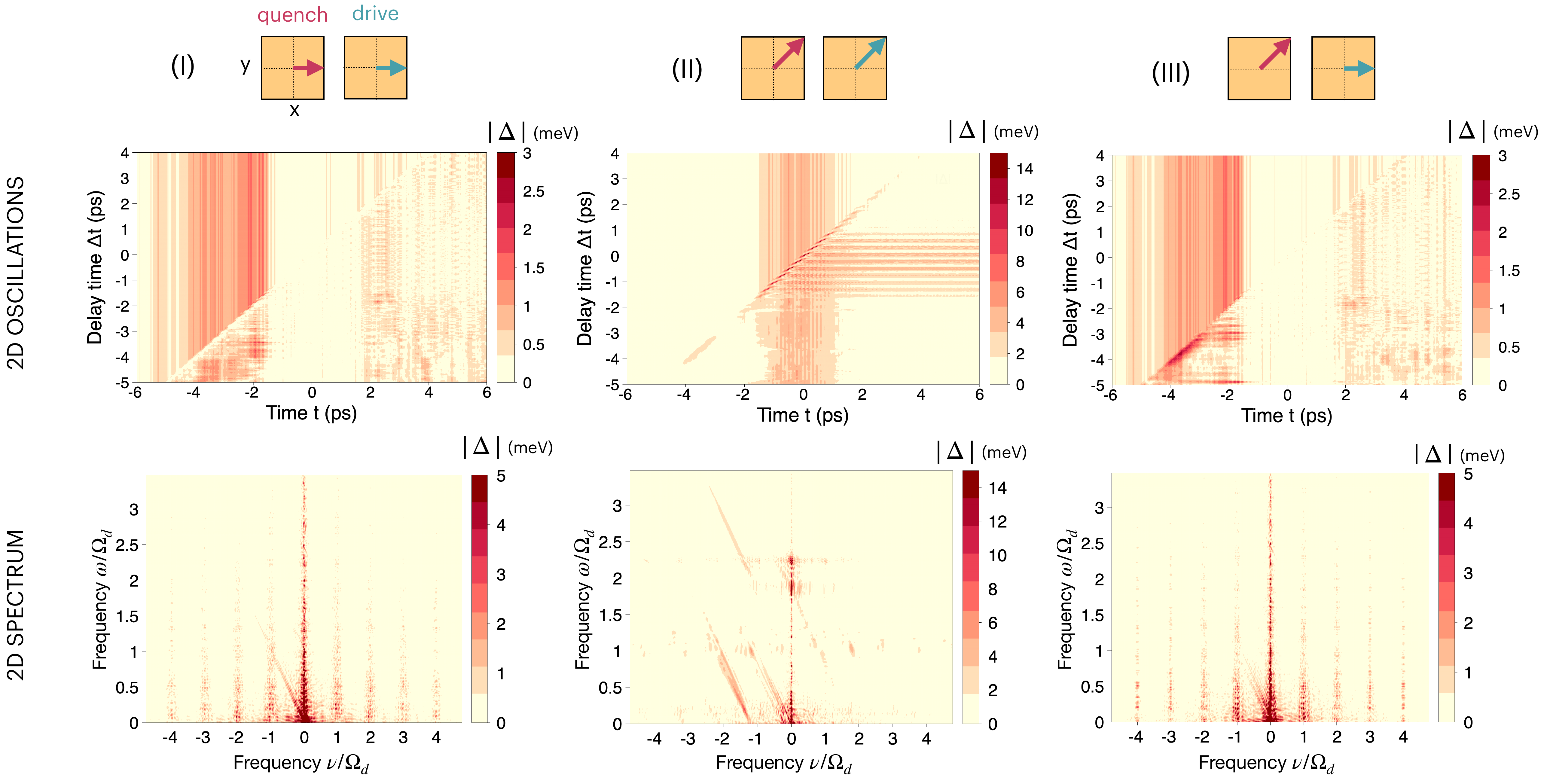}
\caption{Gap oscillations and frequency spectra. (a) 2D oscillations in $(t, \Delta t)$of the absolute value of the superconducting gap, $|\Delta|$, for the three scheme configurations (I)-(III) described in the main text and illustrated by the plots of quench and drive pulses. (b) Absolute value of the 2D Fourier transform of the full complex gap, $| \mathcal{F} \lbrace \Delta  (t, \Delta t) \rbrace| = |\Delta  (\omega, \nu)|$.} \label{Result4}
\end{figure*}

\section{Quench-drive nonlinear response of a superconductor} \label{AppendixQD}

In order to describe a quench-drive experiment we have to choose the appropriate vector potential $\mathbf{A}(t) = \mathbf{A}_{q} (t) + \mathbf{A}_{d} (t) = \mathbf{\overline{A}}_{q} (t-t_q) + \mathbf{\overline{A}}_{d} (t-t_d)$, where $ \mathbf{A}_{q} (t)$ is the quench pulse centered at time $t=t_q$, $ \mathbf{A}_{d} (t)$ is the driving field centered at $t=t_d$. Introducing the time-delay $\Delta t = t_d - t_q$ and putting $t_d=0$ we can rewrite $\mathbf{A}(t) = \mathbf{\overline{A}}_{q} (t + \Delta t) + \mathbf{\overline{A}}_{d} (\overline{t})$. Therefore the expressions in Eq.~(\ref{Blocheqs}) depend on both t and $\Delta t$. \\
The solution of Eq.~(\ref{Blocheqs}) provides the time-dependent pseudospin, from which the time-dependent order parameter $\Delta (t)$ and the generated current $j (t)$ can be calculated. From the self-consistent gap equation we get
\begin{align}
\Delta_{\mathbf{k}} (t) = V f_\mathbf{k}\sum_{\mathbf{k'}} \sigma^x_\mathbf{k'} (t) - i  \sigma^y_\mathbf{k'} (t) \,.
\end{align}
The current generated by the superconductor in this quench-drive setup is given by the expression
\begin{align}
\mathbf{j}(t, \Delta t) = e \sum_{\mathbf{k}} \mathbf{v}_{\mathbf{k} - e \mathbf{A}(t, \Delta t)} \langle \hat{c}^\dagger_{\mathbf{k}, \uparrow} \hat{c}_{\mathbf{k}, \uparrow} +  \hat{c}^\dagger_{\mathbf{k}, \downarrow} \hat{c}_{\mathbf{k}, \downarrow}  \rangle (t, \Delta t)  \,. \label{Fullcurrent1}
\end{align}
In order to separate the linear and the nonlinear contributions to the full generated output current, we first expand the velocity in series of powers of the vector potential $\mathbf{A}$:
\begin{align}
\mathbf{v}_{\mathbf{k} - e \mathbf{A}(t, \Delta t)} = \mathbf{v}_{\mathbf{k}} + \nabla_{\mathbf{A}} \left( \mathbf{v}_{\mathbf{k} - e \mathbf{A}(t, \Delta t)} \right) \big|_{\mathbf{A}= \mathbf{0}} \cdot \mathbf{A} + \dots \,. \label{velocity}
\end{align}
We notice that $\nabla_{\mathbf{A}} (\cdot) \big|_{\mathbf{A} = \mathbf{0}} = \left[ \nabla_{\mathbf{A}} (\mathbf{\kappa}) \ \nabla_{\mathbf{\kappa}} (\cdot) \right] \big|_{\mathbf{A} = \mathbf{0}} =   -e  \ \nabla_{\mathbf{\kappa}} (\cdot) \big|_{\mathbf{A} = \mathbf{0}}$, with $\mathbf{\kappa} = \mathbf{k}- e \mathbf{A}$. Here we omitted the explicit time-dependence of $\mathbf{A}$ and $\mathbf{\kappa}$ from $t$ and $\Delta t$.
Now we can rewrite Eq.~(\ref{velocity}) as
\begin{align}
\mathbf{v}_{\mathbf{\kappa}} = \mathbf{v}_{\mathbf{k}} - e \ \nabla_{\mathbf{\kappa}}  \mathbf{v}_{\mathbf{\kappa}} \big|_{\mathbf{A}= \mathbf{0}} \cdot \mathbf{A} + \dots \,.
\end{align}
In particular, the equivalence $\nabla_{\mathbf{\kappa}} \mathbf{v}_{\mathbf{\kappa}} \big|_{\mathbf{A} = \mathbf{0}} = \nabla_{\mathbf{k}} \mathbf{v}_{\mathbf{k}}$ holds. Therefore we can simplify Eq.~(\ref{velocity}) writing
\begin{align}
\mathbf{v}_{\mathbf{k}} = \mathbf{v}_{\mathbf{k}} - e \ \mathbf{A} \cdot \nabla_{\mathbf{k}}  \mathbf{v}_{\mathbf{k}} + \dots \,.
\end{align}
Additionally, we expand the electron number
\begin{align}
\langle  \hat{n}_{\mathbf{k}} \rangle (t, \Delta t) = 2 z_\mathbf{k} (t, \Delta t) + 2 \hat{\sigma}_\mathbf{k}^z (0) + 1  \,,
\end{align}
where we used the relation $\hat{n}_{\mathbf{k}} = 2 \hat{\sigma}_\mathbf{k}^z + 1$.
Therefore, Eq.~(\ref{Fullcurrent1}) can be expanded in the lowest orders as
\begin{align}
\mathbf{j}(t, \Delta t) \approx e  \sum_{\mathbf{k}} \left( \mathbf{v}_{\mathbf{k}} - e \ \mathbf{A} (t, \Delta t) \cdot \nabla_{\mathbf{k}}  \mathbf{v}_{\mathbf{k}} \right) \nonumber \\
\cdot \left(  2 z_\mathbf{k} (t, \Delta t) + 2 \hat{\sigma}_\mathbf{k}^z (0) + 1 \right) \,. \label{Fullcurrent2}
\end{align}
We can decompose the generated current along a generic $x'$ axis, in order to extract specific symmetry components:
\begin{align}
\mathbf{j}_{x'} (t, \Delta t) & = (\mathbf{j} (t, \Delta t) \cdot \hat{x}' ) \ \hat{x}' \nonumber \\
& = ( j_x (t, \Delta t) \cos{\theta}  + j_y (t, \Delta t) \sin{\theta}) \ \hat{x}' \,.
\end{align}
The contribution to the current in Eq.~(\ref{Fullcurrent2}) at the lowest order in the external field is given by
\begin{align}
\mathbf{j}^{(0)}(t, \Delta t) = e \sum_{\mathbf{k}} \mathbf{v}_\mathbf{k} \left( 2 \hat{\sigma}_\mathbf{k}^z (t, \Delta t) + 1 \right) \,,
\end{align}
which vanishes due to parity. At the next order, the linear term reads
\begin{align}
\mathbf{j}^{(1)}(t, \Delta t) = - e^2 \sum_\mathbf{k} \mathbf{A} (t, \Delta t) \cdot \nabla_{\mathbf{k}}  \mathbf{v}_{\mathbf{k}} \ \left( 2 \hat{\sigma}_\mathbf{k}^z (0) + 1 \right) \,.
\end{align}
 The full nonlinear response, which is given by the sum of all the odd orders of the current expansion, can be conveniently calculated by
\begin{align}
\mathbf{j}^{(NL)} (t, \Delta t) = e \sum_\mathbf{k}  \mathbf{v}_{\mathbf{k} - e \mathbf{A}(t, \Delta t)} (2 \hat{\sigma}^z_\mathbf{k} (t, \Delta t) - 2 \hat{\sigma}^z_\mathbf{k} (0))  \,. \label{NLcurrent2}
\end{align}
We can also explicitly write the expression of the dominant non-vanishing nonlinear term generated by the driving pulse, the third order component, as follows:
\begin{align}
\mathbf{j}^{(3)} (t, \Delta t) = - 2 e^2 \sum_\mathbf{k} z_\mathbf{k} (t, \Delta t) \ \mathbf{A} (t, \Delta t) \cdot \nabla_{\mathbf{k}} \mathbf{v}_{\mathbf{k}} \,, \label{NLcurrent}
\end{align}
where $z_\mathbf{k} (t, \Delta t)$ is the third component of the pseudospin vector $\mathbf{\sigma}_\mathbf{k} (t, \Delta t)$, containing the information of the state of the system perturbed by the quench pulse. The paramagnetic term is neglected here due to suppression by parity \cite{schwarz2020theory}. \\
In general it is useful to extract the 2D frequency spectrum of such a response, in order to analyze the relevant high harmonics: for this reason, we compute the 2D Fourier transform with respect to the evolution time $t$ and the quench-drive delay time $\Delta t$, obtaining the reciprocal variables $\omega \equiv \tilde{\mathcal{F}} (t)$ and $\nu \equiv \tilde{\mathcal{F}} (\Delta t)$, respectively. \\
As an example, the 2D Fourier transform of Eq.(\ref{NLcurrent}) to provide the third harmonic response of the driving frequency along the direction $\hat{x'}$ is
\begin{align}
j_{x'}^{(3)} (\omega = 3 \Omega_d, \nu) &= -2 \text{e}^2 A_d \sum_\mathbf{k} F_{x'}(\mathbf{k}) \ z_\mathbf{k} (\omega = 2 \Omega_d, \nu) \notag \\
 &- 2 \text{e}^2 A_d \sum_\mathbf{k} F_{x'} (\mathbf{k}) \ z_\mathbf{k} (\omega = 4 \Omega_d, \nu) \,,
\end{align}
where $F_{x'}$ is an appropriate function independent of the frequency which contains information on the driving shape, the measurement axis and the quasiparticles' momentum.

\begin{figure*}[tb!]
\centering
\includegraphics[width=18cm]{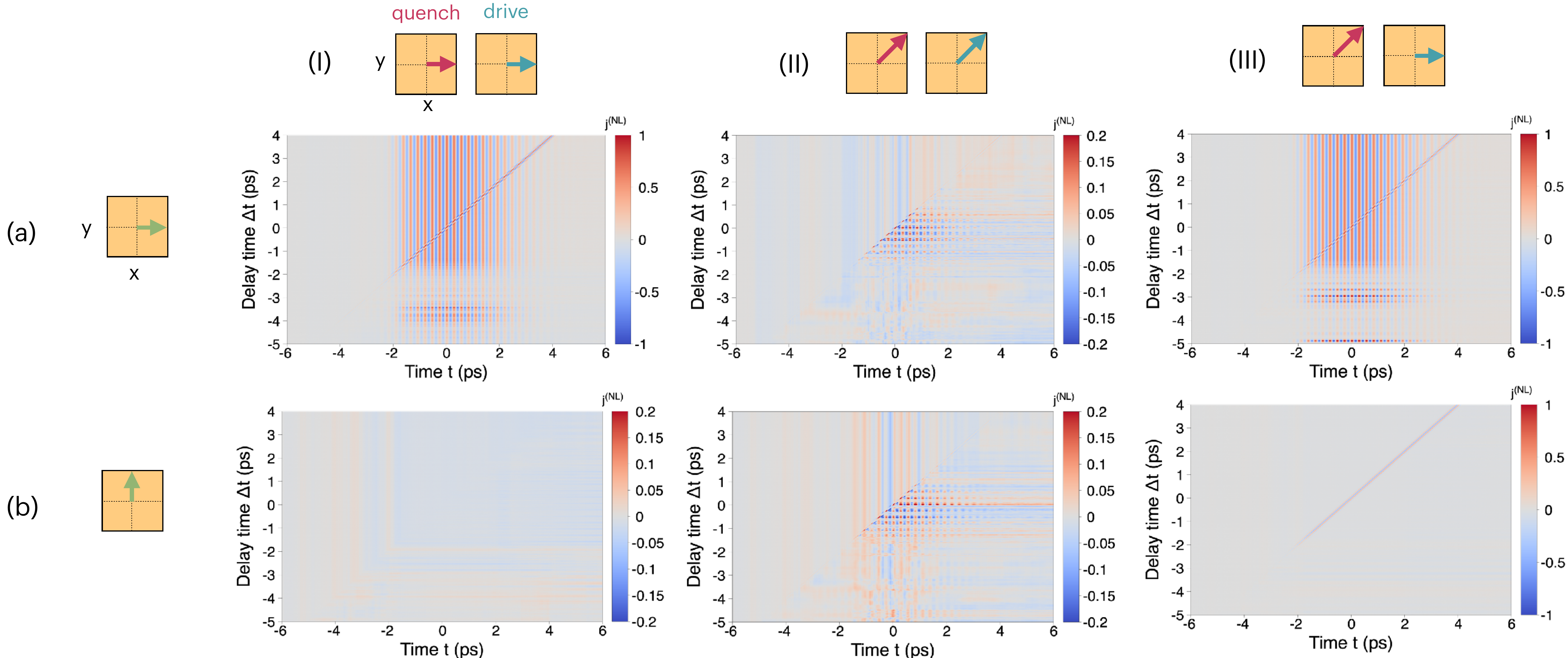}
\caption{2D nonlinear current. Plots of the generated nonlinear current as a function of real time $t$ and the quench-drive delay time $\Delta t$, for three different schemes (I)-(III) described in the main text, and two polarized output measures along (a) $x$ and (b) $y$ axis, respectively. This figure corresponds to Fig.~\ref{Result2}, here obtained with different frequencies of quench and drive pulses, as explained in the main text. Be aware of the different color scale for each plot and with respect to Fig.~\ref{Result2}.} \label{Result5}
\end{figure*}

\begin{figure*}[tb!]
\centering
\includegraphics[width=18cm]{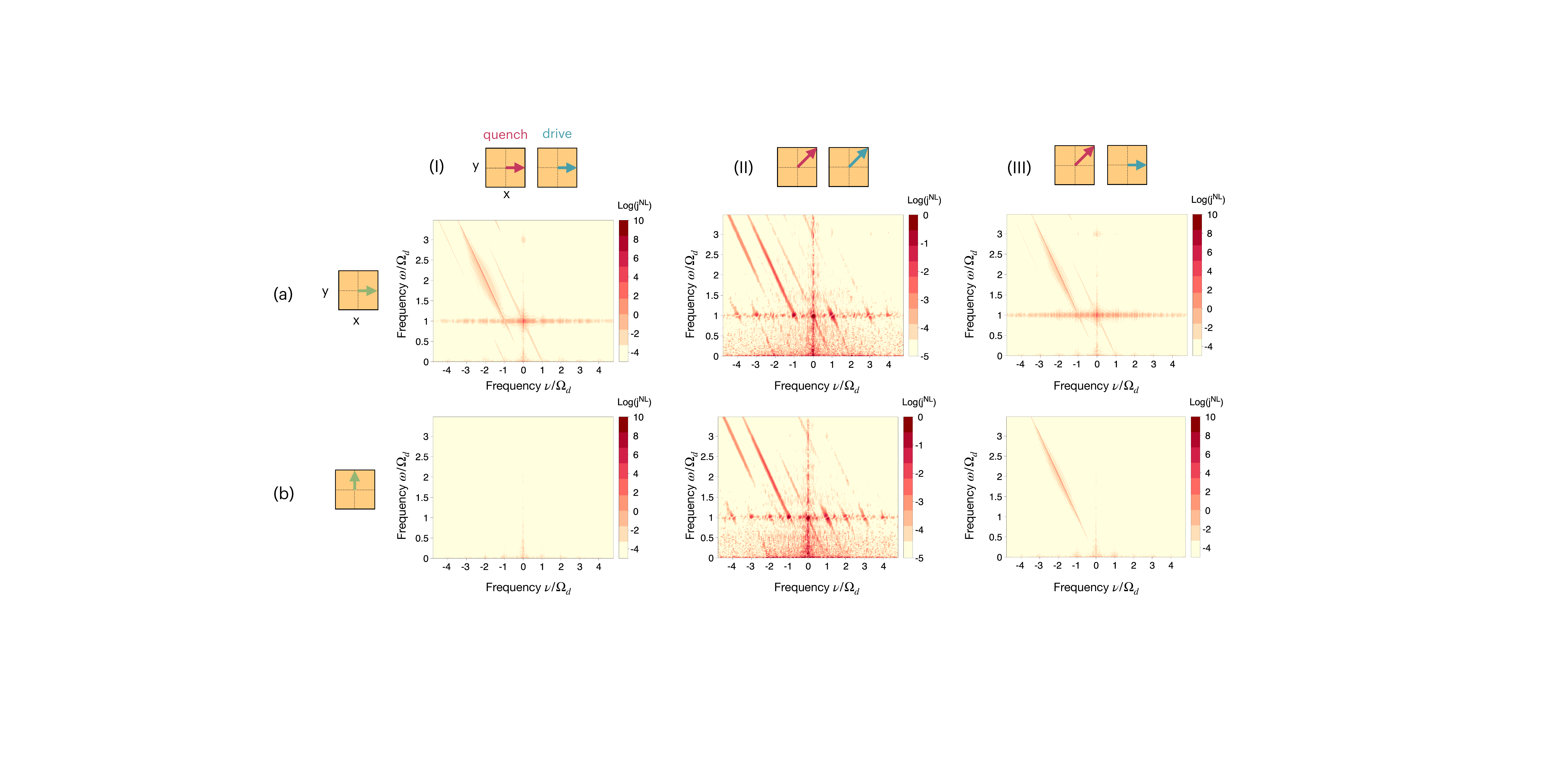}
\caption{2D nonlinear current spectra. Plots of the Fourier transform of the nonlinear current in Fig.\ref{Result5}, for three different schemes (I)-(III) described in the main text, and two polarized output measures along (a) $x$ and (b) $y$ axis, respectively. This figure corresponds to Fig.\ref{Result3}, here obtained with different frequencies of quench and drive pulses, as explained in the main text. Be aware of the different Log color scale for each plot. } \label{Result6}
\end{figure*}

\begin{figure*}[h!]
\centering
\includegraphics[width=18cm]{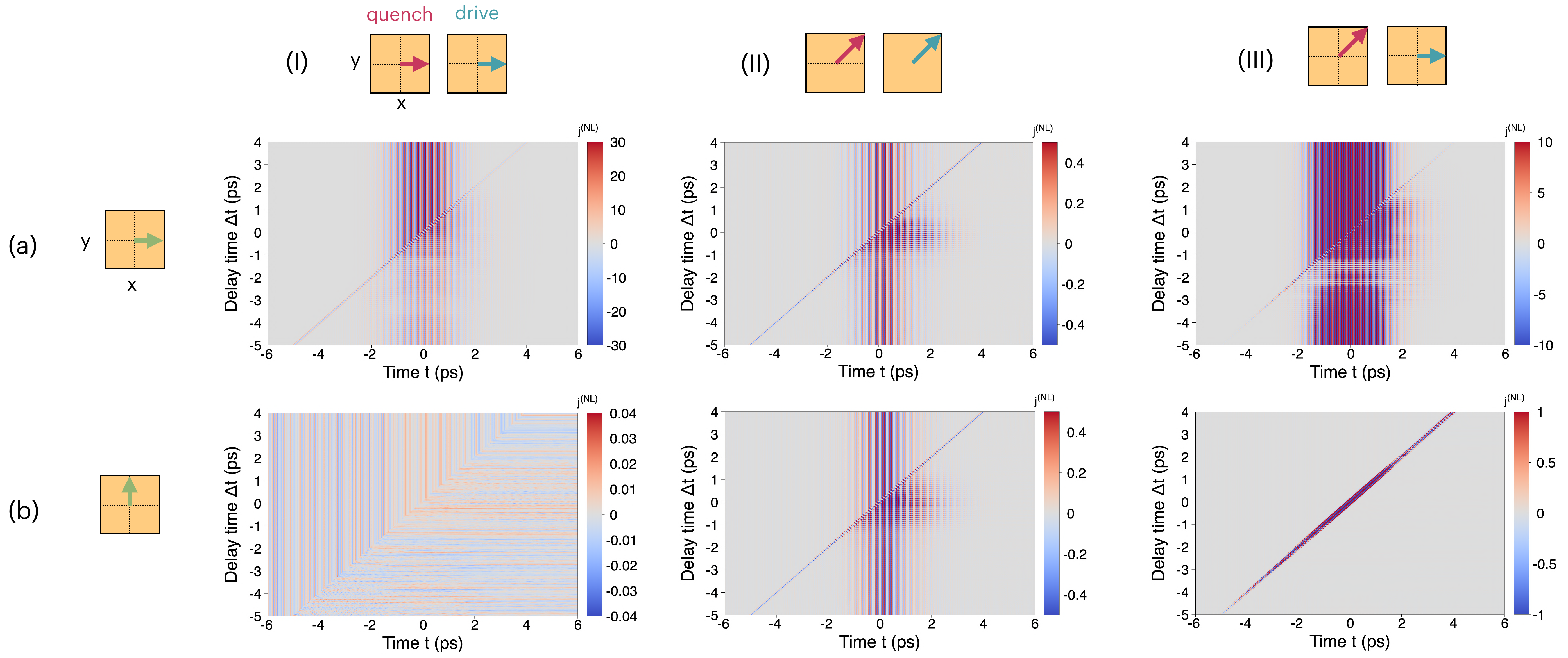}
\caption{2D nonlinear current for partially incoherent pairs. Plots of the generated nonlinear current as a function of real time $t$ and the quench-drive delay time $\Delta t$, for three different schemes (I)-(III) described in the main text, and two polarized output measures along (a) $x$ and (b) $y$ axis, respectively. This figure corresponds to Fig.~\ref{Result2}, here obtained for phase-fluctuating superconductors with $\phi_{max} = \pi/8$. Be aware of the different color scale for each plot and with respect to Fig.~\ref{Result2}.} \label{Result7}
\end{figure*}

\begin{figure*}[h!]
\centering
\includegraphics[width=18cm]{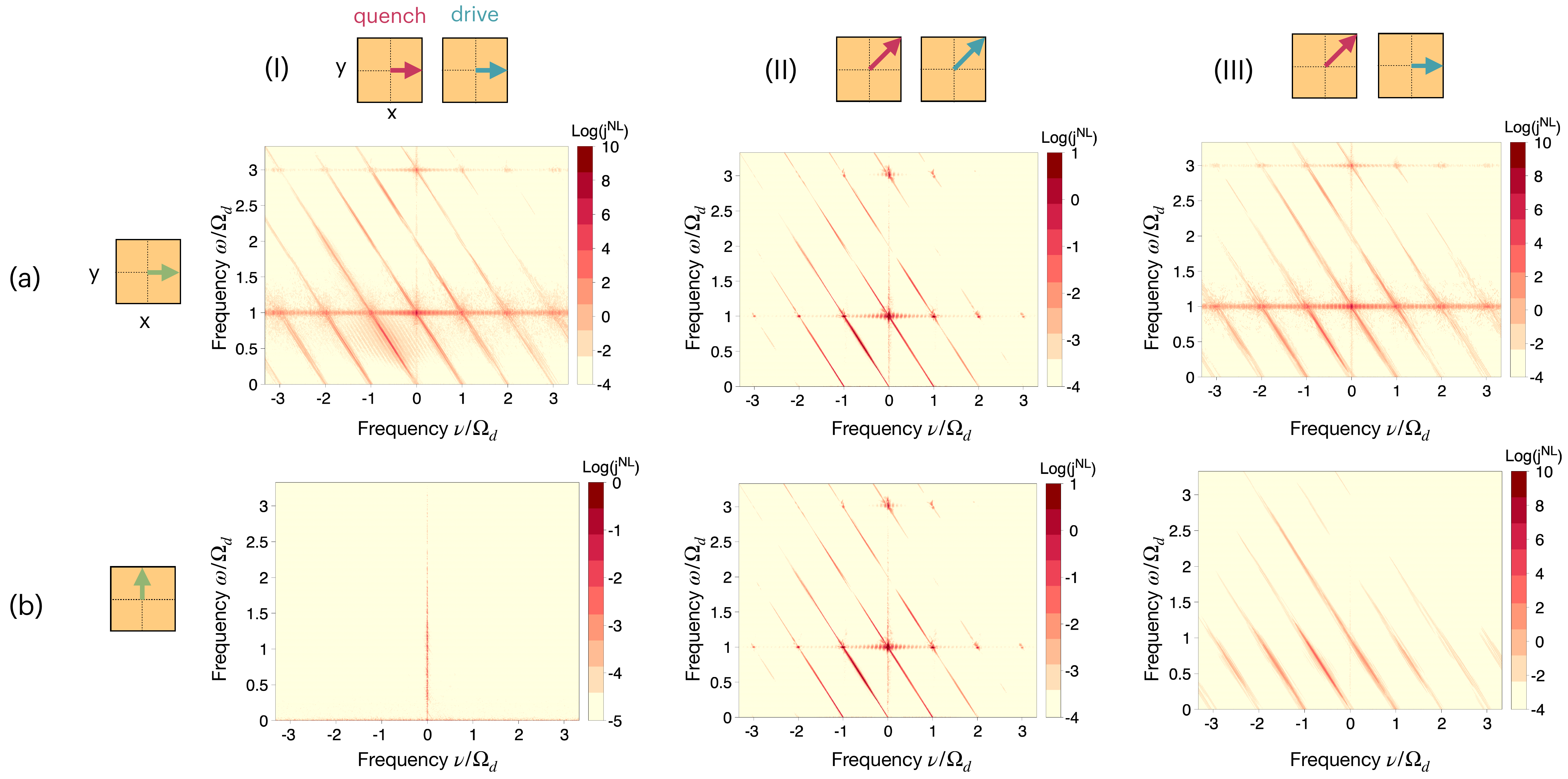}
\caption{2D nonlinear current spectra for partially incoherent pairs. Plots of the Fourier transform of the nonlinear current in Fig.\ref{Result5}, for three different schemes (I)-(III) described in the main text, and two polarized output measures along (a) $x$ and (b) $y$ axis, respectively. This figure corresponds to Fig.\ref{Result3}, here obtained for phase-fluctuating superconductors with $\phi_{max} = \pi/8$. Be aware of the different Log color scale for each plot.} \label{Result8}
\end{figure*}

\section{Additional results} \label{MoreResults}
We present here additional results, obtained for the same quench and drive intensities as the ones in the main text, i.e. $A_d = A_q = 0.8$, as well as pulses' duration and shape, but with different frequencies: namely, $\Omega_d = 3.66$ THz and $\Omega_q = 7.48$ THz, respectively. As a consequence, the quench is nearly resonant with the maximum superconducting equilibrium gap, $\Delta_{max} = 31$ meV $= 7.5$ THz, while the driving pulse is far from it. The generated nonlinear current is therefore affected by these conditions, and the response appears in some cases qualitatively and quantitatively different from the one obtained in the main text, even if the symmetries involved in the quench-drive spectra are the same. \\
\indent We first analyze the behavior of the absolute value of the superconducting gap as a function of the real time $t$ and the delay time $\Delta t$. Interestingly, we realize that the gap is very poorly excited in configurations (I) and (III) due to the $B_{1g}$ symmetry and the driving contribution, with a maximum amplitude of about $3$ meV, and as low as $1$ meV on the central peak of the driving field at $t =0$. On the other hand, the scheme (II) has a higher gap excitation. 
The corresponding 2D Fourier spectra show that, for schemes (I) and (III) there are no proper gap oscillations, but rather an almost frequency-independent enhancement, plus quench-induced contributions (vertical lines in Fig.~\ref{Result4}(I),(III)(b)). On the other hand, for the scheme (II) with diagonal quench and drive pulses, where the $B_{2g}$ symmetry is excited, a gap oscillation at $\omega = 2 \Omega_d$ due to the driving appears, as well as at $\omega = 2 \Delta$, which includes Higgs and quasiparticles' excitations at twice the induced gap amplitude, around $8.5$ meV.  \\
\indent We now turn to the generated current: due to the different resonance conditions, we expect the current responses involving a quench pulse to be more intense, saturating the purely drive signals. In particular, for schemes (I) and (III) where the gap excitation and oscillations are much smaller, we expect the susceptibility term independent of the frequency to be the most relevant \cite{Cea2016nonlinear}. \\
In Fig.s \ref{Result5} and \ref{Result6} the measured nonlinear responses in time and the corresponding 2D Fourier spectra are shown, respectively. We notice that the current measured along the $x$ axis for schemes (I) and (III), involving mainly the $B_{1g}$ symmetry, is quantitatively different from the one in Fig.~\ref{Result2}. The lower intensity (the scales of Fig.s \ref{Result2} and \ref{Result5} are different), in fact, is explained by the fact that the main response involves the driving pulse, and the corresponding susceptibility is now more far from resonance. On the other hand, the scheme (II), with quench and drive pulses along the $\hat{xy}$ diagonal axis, provides now a slightly stronger response, involving mainly the quench pulse. \\
\indent The 2D Fourier spectra in Fig.~\ref{Result6} are even more dense of information. In fact, the spectra of schemes (I) and (III) present much fewer features than with the choices of frequency in the main text: in particular, (I)(a) and (II)(a) have a weaker third harmonic generation and only one diagonal line, representing the nonequilibrium modulation due to the quench pulses. Moreover, the current measured along the $y$ axis in (III)(b) has no first harmonic contribution, and is saturated by the same nonequilibrium modulation of (I),(III) (a). On the other hand, the spectra of (II) are much more complex, exhibiting more and stronger frequency modulations and the emergence of a non-equilibrium third harmonic at $\omega = 3 \Omega_d, \nu = 0$. \\
All in all, we have observed how the nonlinear signal is still present in (I) and (III) configurations, the current intensity being higher than in scheme (II), even if the gap is less excited in the former. The reason of this behavior can be ascribed once again to the symmetries involved and here identified. \\
We now turn to consider a phase-fluctuating superconductor with $\phi_{max} = \pi/8$. In this situation, the superconducting order parameter is nonzero even in equilibrium, and the resonance energy of quasiparticles' fluctuations is much higher. We present in Fig.~\ref{Result7}, \ref{Result8} the symmetry-resolved 2D nonlinear current and the corresponding Fourier transform signals in frequency, respectively. As we can notice, these are qualitatively similar to the results of Fig.~\ref{Result2}, \ref{Result3}: this demonstrates that the nonlinear current response originates mainly from the presence of the underlying Cooper pairing, which allows to transiently enhance the superconducting gap in the quench-drive setup. However, we notice here significant quantitative differences with respect to the purely incoherent case. In fact, it is possible to notice in Fig.~\ref{Result7} the higher intensity in all the non-vanishing responses. Interestingly, this quantitative aspect leads to a non-negligible qualitative difference: namely, the third harmonic signal for configurations II(a),(b) [Fig.~\ref{Result8}] is quantitatively distinguishable, and the one in I(a) and III(a) has numerous features which modulates its time-evolution with frequencies multiple of the driving frequency $\Omega_d$. In conclusion, we can claim that the main features of the symmetry-resolved nonlinear quench-drive spectroscopy are determined by the symmetry of the underlying Cooper pairing, allowing to distinguish whether there are pre-formed pairs even in absence of a superconducting order parameter. However, it would be possible to discriminate between a fully-incoherent and a partially-incoherent superconductor by analysing the intensity of the third harmonic signal and especially at its time- and frequency-dependent modulations for given symmetries.


\begin{thebibliography}{58}%
\makeatletter
\providecommand \@ifxundefined [1]{%
 \@ifx{#1\undefined}
}%
\providecommand \@ifnum [1]{%
 \ifnum #1\expandafter \@firstoftwo
 \else \expandafter \@secondoftwo
 \fi
}%
\providecommand \@ifx [1]{%
 \ifx #1\expandafter \@firstoftwo
 \else \expandafter \@secondoftwo
 \fi
}%
\providecommand \natexlab [1]{#1}%
\providecommand \enquote  [1]{``#1''}%
\providecommand \bibnamefont  [1]{#1}%
\providecommand \bibfnamefont [1]{#1}%
\providecommand \citenamefont [1]{#1}%
\providecommand \href@noop [0]{\@secondoftwo}%
\providecommand \href [0]{\begingroup \@sanitize@url \@href}%
\providecommand \@href[1]{\@@startlink{#1}\@@href}%
\providecommand \@@href[1]{\endgroup#1\@@endlink}%
\providecommand \@sanitize@url [0]{\catcode `\\12\catcode `\$12\catcode
  `\&12\catcode `\#12\catcode `\^12\catcode `\_12\catcode `\%12\relax}%
\providecommand \@@startlink[1]{}%
\providecommand \@@endlink[0]{}%
\providecommand \url  [0]{\begingroup\@sanitize@url \@url }%
\providecommand \@url [1]{\endgroup\@href {#1}{\urlprefix }}%
\providecommand \urlprefix  [0]{URL }%
\providecommand \Eprint [0]{\href }%
\providecommand \doibase [0]{https://doi.org/}%
\providecommand \selectlanguage [0]{\@gobble}%
\providecommand \bibinfo  [0]{\@secondoftwo}%
\providecommand \bibfield  [0]{\@secondoftwo}%
\providecommand \translation [1]{[#1]}%
\providecommand \BibitemOpen [0]{}%
\providecommand \bibitemStop [0]{}%
\providecommand \bibitemNoStop [0]{.\EOS\space}%
\providecommand \EOS [0]{\spacefactor3000\relax}%
\providecommand \BibitemShut  [1]{\csname bibitem#1\endcsname}%
\let\auto@bib@innerbib\@empty
\bibitem [{\citenamefont {Wu}\ \emph {et~al.}(1987)\citenamefont {Wu},
  \citenamefont {Ashburn}, \citenamefont {Torng}, \citenamefont {Hor},
  \citenamefont {Meng}, \citenamefont {Gao}, \citenamefont {Huang},
  \citenamefont {Wang},\ and\ \citenamefont {Chu}}]{Wu1987}%
  \BibitemOpen
  \bibfield  {author} {\bibinfo {author} {\bibfnamefont {M.~K.}\ \bibnamefont
  {Wu}}, \bibinfo {author} {\bibfnamefont {J.~R.}\ \bibnamefont {Ashburn}},
  \bibinfo {author} {\bibfnamefont {C.~J.}\ \bibnamefont {Torng}}, \bibinfo
  {author} {\bibfnamefont {P.~H.}\ \bibnamefont {Hor}}, \bibinfo {author}
  {\bibfnamefont {R.~L.}\ \bibnamefont {Meng}}, \bibinfo {author}
  {\bibfnamefont {L.}~\bibnamefont {Gao}}, \bibinfo {author} {\bibfnamefont
  {Z.~J.}\ \bibnamefont {Huang}}, \bibinfo {author} {\bibfnamefont {Y.~Q.}\
  \bibnamefont {Wang}},\ and\ \bibinfo {author} {\bibfnamefont {C.~W.}\
  \bibnamefont {Chu}},\ }\bibfield  {title} {\bibinfo {title}
  {Superconductivity at 93 k in a new mixed-phase y-ba-cu-o compound system at
  ambient pressure},\ }\href {https://doi.org/10.1103/PhysRevLett.58.908}
  {\bibfield  {journal} {\bibinfo  {journal} {Phys. Rev. Lett.}\ }\textbf
  {\bibinfo {volume} {58}},\ \bibinfo {pages} {908} (\bibinfo {year}
  {1987})}\BibitemShut {NoStop}%
\bibitem [{\citenamefont {Maeda}\ \emph {et~al.}(1988)\citenamefont {Maeda},
  \citenamefont {Tanaka}, \citenamefont {Fukutomi},\ and\ \citenamefont
  {Asano}}]{Maeda_1988}%
  \BibitemOpen
  \bibfield  {author} {\bibinfo {author} {\bibfnamefont {H.}~\bibnamefont
  {Maeda}}, \bibinfo {author} {\bibfnamefont {Y.}~\bibnamefont {Tanaka}},
  \bibinfo {author} {\bibfnamefont {M.}~\bibnamefont {Fukutomi}},\ and\
  \bibinfo {author} {\bibfnamefont {T.}~\bibnamefont {Asano}},\ }\bibfield
  {title} {\bibinfo {title} {A new high-tc oxide superconductor without a rare
  earth element},\ }\href {https://doi.org/10.1143/JJAP.27.L209} {\bibfield
  {journal} {\bibinfo  {journal} {Japanese Journal of Applied Physics}\
  }\textbf {\bibinfo {volume} {27}},\ \bibinfo {pages} {L209} (\bibinfo {year}
  {1988})}\BibitemShut {NoStop}%
\bibitem [{\citenamefont {Keimer}\ \emph {et~al.}(2015)\citenamefont {Keimer},
  \citenamefont {Kivelson}, \citenamefont {Norman}, \citenamefont {Uchida},\
  and\ \citenamefont {Zaanen}}]{Keimer2015}%
  \BibitemOpen
  \bibfield  {author} {\bibinfo {author} {\bibfnamefont {B.}~\bibnamefont
  {Keimer}}, \bibinfo {author} {\bibfnamefont {S.~A.}\ \bibnamefont
  {Kivelson}}, \bibinfo {author} {\bibfnamefont {M.~R.}\ \bibnamefont
  {Norman}}, \bibinfo {author} {\bibfnamefont {S.}~\bibnamefont {Uchida}},\
  and\ \bibinfo {author} {\bibfnamefont {J.}~\bibnamefont {Zaanen}},\
  }\bibfield  {title} {\bibinfo {title} {From quantum matter to
  high-temperature superconductivity in copper oxides},\ }\href
  {https://doi.org/10.1038/nature14165} {\bibfield  {journal} {\bibinfo
  {journal} {Nature}\ }\textbf {\bibinfo {volume} {518}},\ \bibinfo {pages}
  {179} (\bibinfo {year} {2015})}\BibitemShut {NoStop}%
\bibitem [{\citenamefont {Toda}\ \emph {et~al.}(2014)\citenamefont {Toda},
  \citenamefont {Kawanokami}, \citenamefont {Kurosawa}, \citenamefont {Oda},
  \citenamefont {Madan}, \citenamefont {Mertelj}, \citenamefont {Kabanov},\
  and\ \citenamefont {Mihailovic}}]{Toda2014rotational}%
  \BibitemOpen
  \bibfield  {author} {\bibinfo {author} {\bibfnamefont {Y.}~\bibnamefont
  {Toda}}, \bibinfo {author} {\bibfnamefont {F.}~\bibnamefont {Kawanokami}},
  \bibinfo {author} {\bibfnamefont {T.}~\bibnamefont {Kurosawa}}, \bibinfo
  {author} {\bibfnamefont {M.}~\bibnamefont {Oda}}, \bibinfo {author}
  {\bibfnamefont {I.}~\bibnamefont {Madan}}, \bibinfo {author} {\bibfnamefont
  {T.}~\bibnamefont {Mertelj}}, \bibinfo {author} {\bibfnamefont {V.~V.}\
  \bibnamefont {Kabanov}},\ and\ \bibinfo {author} {\bibfnamefont
  {D.}~\bibnamefont {Mihailovic}},\ }\bibfield  {title} {\bibinfo {title}
  {Rotational symmetry breaking in
  ${\mathrm{bi}}_{2}{\mathrm{sr}}_{2}{\mathrm{cacu}}_{2}{\mathrm{o}}_{8+\ensuremath{\delta}}$
  probed by polarized femtosecond spectroscopy},\ }\href
  {https://doi.org/10.1103/PhysRevB.90.094513} {\bibfield  {journal} {\bibinfo
  {journal} {Phys. Rev. B}\ }\textbf {\bibinfo {volume} {90}},\ \bibinfo
  {pages} {094513} (\bibinfo {year} {2014})}\BibitemShut {NoStop}%
\bibitem [{\citenamefont {Devereaux}\ \emph {et~al.}(1994)\citenamefont
  {Devereaux}, \citenamefont {Einzel}, \citenamefont {Stadlober}, \citenamefont
  {Hackl}, \citenamefont {Leach},\ and\ \citenamefont
  {Neumeier}}]{Devereaux1994Electronic}%
  \BibitemOpen
  \bibfield  {author} {\bibinfo {author} {\bibfnamefont {T.~P.}\ \bibnamefont
  {Devereaux}}, \bibinfo {author} {\bibfnamefont {D.}~\bibnamefont {Einzel}},
  \bibinfo {author} {\bibfnamefont {B.}~\bibnamefont {Stadlober}}, \bibinfo
  {author} {\bibfnamefont {R.}~\bibnamefont {Hackl}}, \bibinfo {author}
  {\bibfnamefont {D.~H.}\ \bibnamefont {Leach}},\ and\ \bibinfo {author}
  {\bibfnamefont {J.~J.}\ \bibnamefont {Neumeier}},\ }\bibfield  {title}
  {\bibinfo {title} {{Electronic Raman scattering in
  high-${\mathit{T}}_{\mathit{c}}$ superconductors: A probe of
  ${\mathit{d}}_{\mathit{x}}^{2}$-${\mathit{y}}^{2}$ pairing}},\ }\href
  {https://doi.org/10.1103/PhysRevLett.72.396} {\bibfield  {journal} {\bibinfo
  {journal} {Phys. Rev. Lett.}\ }\textbf {\bibinfo {volume} {72}},\ \bibinfo
  {pages} {396} (\bibinfo {year} {1994})}\BibitemShut {NoStop}%
\bibitem [{\citenamefont {Markiewicz}\ \emph {et~al.}(2005)\citenamefont
  {Markiewicz}, \citenamefont {Sahrakorpi}, \citenamefont {Lindroos},
  \citenamefont {Lin},\ and\ \citenamefont {Bansil}}]{Markiewicz2005}%
  \BibitemOpen
  \bibfield  {author} {\bibinfo {author} {\bibfnamefont {R.~S.}\ \bibnamefont
  {Markiewicz}}, \bibinfo {author} {\bibfnamefont {S.}~\bibnamefont
  {Sahrakorpi}}, \bibinfo {author} {\bibfnamefont {M.}~\bibnamefont
  {Lindroos}}, \bibinfo {author} {\bibfnamefont {H.}~\bibnamefont {Lin}},\ and\
  \bibinfo {author} {\bibfnamefont {A.}~\bibnamefont {Bansil}},\ }\bibfield
  {title} {\bibinfo {title} {One-band tight-binding model parametrization of
  the high-${T}_{c}$ cuprates including the effect of ${k}_{z}$ dispersion},\
  }\href {https://doi.org/10.1103/PhysRevB.72.054519} {\bibfield  {journal}
  {\bibinfo  {journal} {Phys. Rev. B}\ }\textbf {\bibinfo {volume} {72}},\
  \bibinfo {pages} {054519} (\bibinfo {year} {2005})}\BibitemShut {NoStop}%
\bibitem [{\citenamefont {Matsunaga}\ \emph {et~al.}(2014)\citenamefont
  {Matsunaga}, \citenamefont {Tsuji}, \citenamefont {Fujita}, \citenamefont
  {Sugioka}, \citenamefont {Makise}, \citenamefont {Uzawa}, \citenamefont
  {Terai}, \citenamefont {Wang}, \citenamefont {Aoki},\ and\ \citenamefont
  {Shimano}}]{Matsunaga2014}%
  \BibitemOpen
  \bibfield  {author} {\bibinfo {author} {\bibfnamefont {R.}~\bibnamefont
  {Matsunaga}}, \bibinfo {author} {\bibfnamefont {N.}~\bibnamefont {Tsuji}},
  \bibinfo {author} {\bibfnamefont {H.}~\bibnamefont {Fujita}}, \bibinfo
  {author} {\bibfnamefont {A.}~\bibnamefont {Sugioka}}, \bibinfo {author}
  {\bibfnamefont {K.}~\bibnamefont {Makise}}, \bibinfo {author} {\bibfnamefont
  {Y.}~\bibnamefont {Uzawa}}, \bibinfo {author} {\bibfnamefont
  {H.}~\bibnamefont {Terai}}, \bibinfo {author} {\bibfnamefont
  {Z.}~\bibnamefont {Wang}}, \bibinfo {author} {\bibfnamefont {H.}~\bibnamefont
  {Aoki}},\ and\ \bibinfo {author} {\bibfnamefont {R.}~\bibnamefont
  {Shimano}},\ }\bibfield  {title} {\bibinfo {title} {Light-induced collective
  pseudospin precession resonating with higgs mode in a superconductor},\
  }\href {https://doi.org/10.1126/science.1254697} {\bibfield  {journal}
  {\bibinfo  {journal} {Science}\ }\textbf {\bibinfo {volume} {345}},\ \bibinfo
  {pages} {1145–1149} (\bibinfo {year} {2014})}\BibitemShut {NoStop}%
\bibitem [{\citenamefont {Cea}\ \emph {et~al.}(2016)\citenamefont {Cea},
  \citenamefont {Castellani},\ and\ \citenamefont
  {Benfatto}}]{Cea2016nonlinear}%
  \BibitemOpen
  \bibfield  {author} {\bibinfo {author} {\bibfnamefont {T.}~\bibnamefont
  {Cea}}, \bibinfo {author} {\bibfnamefont {C.}~\bibnamefont {Castellani}},\
  and\ \bibinfo {author} {\bibfnamefont {L.}~\bibnamefont {Benfatto}},\
  }\bibfield  {title} {\bibinfo {title} {{Nonlinear optical effects and
  third-harmonic generation in superconductors: Cooper pairs versus Higgs mode
  contribution}},\ }\href {https://doi.org/10.1103/PhysRevB.93.180507}
  {\bibfield  {journal} {\bibinfo  {journal} {Phys. Rev. B}\ }\textbf {\bibinfo
  {volume} {93}},\ \bibinfo {pages} {180507} (\bibinfo {year}
  {2016})}\BibitemShut {NoStop}%
\bibitem [{\citenamefont {Katsumi}\ \emph {et~al.}(2018)\citenamefont
  {Katsumi}, \citenamefont {Tsuji}, \citenamefont {Hamada}, \citenamefont
  {Matsunaga}, \citenamefont {Schneeloch}, \citenamefont {Zhong}, \citenamefont
  {Gu}, \citenamefont {Aoki}, \citenamefont {Gallais},\ and\ \citenamefont
  {Shimano}}]{Katsumi2018higgs}%
  \BibitemOpen
  \bibfield  {author} {\bibinfo {author} {\bibfnamefont {K.}~\bibnamefont
  {Katsumi}}, \bibinfo {author} {\bibfnamefont {N.}~\bibnamefont {Tsuji}},
  \bibinfo {author} {\bibfnamefont {Y.~I.}\ \bibnamefont {Hamada}}, \bibinfo
  {author} {\bibfnamefont {R.}~\bibnamefont {Matsunaga}}, \bibinfo {author}
  {\bibfnamefont {J.}~\bibnamefont {Schneeloch}}, \bibinfo {author}
  {\bibfnamefont {R.~D.}\ \bibnamefont {Zhong}}, \bibinfo {author}
  {\bibfnamefont {G.~D.}\ \bibnamefont {Gu}}, \bibinfo {author} {\bibfnamefont
  {H.}~\bibnamefont {Aoki}}, \bibinfo {author} {\bibfnamefont {Y.}~\bibnamefont
  {Gallais}},\ and\ \bibinfo {author} {\bibfnamefont {R.}~\bibnamefont
  {Shimano}},\ }\bibfield  {title} {\bibinfo {title} {Higgs mode in the
  $d$-wave superconductor
  ${\mathrm{bi}}_{2}{\mathrm{sr}}_{2}{\mathrm{cacu}}_{2}{\mathrm{o}}_{8+x}$
  driven by an intense terahertz pulse},\ }\href
  {https://doi.org/10.1103/PhysRevLett.120.117001} {\bibfield  {journal}
  {\bibinfo  {journal} {Phys. Rev. Lett.}\ }\textbf {\bibinfo {volume} {120}},\
  \bibinfo {pages} {117001} (\bibinfo {year} {2018})}\BibitemShut {NoStop}%
\bibitem [{\citenamefont {Chu}\ \emph {et~al.}(2020)\citenamefont {Chu},
  \citenamefont {Kim}, \citenamefont {Katsumi}, \citenamefont {Kovalev},
  \citenamefont {Dawson}, \citenamefont {Schwarz}, \citenamefont {Yoshikawa},
  \citenamefont {Kim}, \citenamefont {Putzky}, \citenamefont {Li} \emph
  {et~al.}}]{chu2020phase}%
  \BibitemOpen
  \bibfield  {author} {\bibinfo {author} {\bibfnamefont {H.}~\bibnamefont
  {Chu}}, \bibinfo {author} {\bibfnamefont {M.-J.}\ \bibnamefont {Kim}},
  \bibinfo {author} {\bibfnamefont {K.}~\bibnamefont {Katsumi}}, \bibinfo
  {author} {\bibfnamefont {S.}~\bibnamefont {Kovalev}}, \bibinfo {author}
  {\bibfnamefont {R.~D.}\ \bibnamefont {Dawson}}, \bibinfo {author}
  {\bibfnamefont {L.}~\bibnamefont {Schwarz}}, \bibinfo {author} {\bibfnamefont
  {N.}~\bibnamefont {Yoshikawa}}, \bibinfo {author} {\bibfnamefont
  {G.}~\bibnamefont {Kim}}, \bibinfo {author} {\bibfnamefont {D.}~\bibnamefont
  {Putzky}}, \bibinfo {author} {\bibfnamefont {Z.~Z.}\ \bibnamefont {Li}},
  \emph {et~al.},\ }\bibfield  {title} {\bibinfo {title} {Phase-resolved higgs
  response in superconducting cuprates},\ }\href@noop {} {\bibfield  {journal}
  {\bibinfo  {journal} {Nature communications}\ }\textbf {\bibinfo {volume}
  {11}},\ \bibinfo {pages} {1} (\bibinfo {year} {2020})}\BibitemShut {NoStop}%
\bibitem [{\citenamefont {Puviani}\ \emph {et~al.}(2021)\citenamefont
  {Puviani}, \citenamefont {Baum}, \citenamefont {Ono}, \citenamefont {Ando},
  \citenamefont {Hackl},\ and\ \citenamefont {Manske}}]{Puviani2021PRL}%
  \BibitemOpen
  \bibfield  {author} {\bibinfo {author} {\bibfnamefont {M.}~\bibnamefont
  {Puviani}}, \bibinfo {author} {\bibfnamefont {A.}~\bibnamefont {Baum}},
  \bibinfo {author} {\bibfnamefont {S.}~\bibnamefont {Ono}}, \bibinfo {author}
  {\bibfnamefont {Y.}~\bibnamefont {Ando}}, \bibinfo {author} {\bibfnamefont
  {R.}~\bibnamefont {Hackl}},\ and\ \bibinfo {author} {\bibfnamefont
  {D.}~\bibnamefont {Manske}},\ }\bibfield  {title} {\bibinfo {title}
  {Calculation of an enhanced ${A}_{1g}$ symmetry mode induced by higgs
  oscillations in the raman spectrum of high-temperature cuprate
  superconductors},\ }\href {https://doi.org/10.1103/PhysRevLett.127.197001}
  {\bibfield  {journal} {\bibinfo  {journal} {Phys. Rev. Lett.}\ }\textbf
  {\bibinfo {volume} {127}},\ \bibinfo {pages} {197001} (\bibinfo {year}
  {2021})}\BibitemShut {NoStop}%
\bibitem [{\citenamefont {Chu}\ \emph {et~al.}(2021)\citenamefont {Chu},
  \citenamefont {Kovalev}, \citenamefont {Wang}, \citenamefont {Schwarz},
  \citenamefont {Dong}, \citenamefont {Feng}, \citenamefont {Haenel},
  \citenamefont {Kim}, \citenamefont {Shabestari}, \citenamefont {Phuong},
  \citenamefont {Honasoge}, \citenamefont {Dawson}, \citenamefont {Putzky},
  \citenamefont {Kim}, \citenamefont {Puviani}, \citenamefont {Chen},
  \citenamefont {Awari}, \citenamefont {Ponomaryov}, \citenamefont {Ilyakov},
  \citenamefont {Bluschke}, \citenamefont {Boschini}, \citenamefont {Zonno},
  \citenamefont {Zhdanovich}, \citenamefont {Na}, \citenamefont {Christiani},
  \citenamefont {Logvenov}, \citenamefont {Jones}, \citenamefont {Damascelli},
  \citenamefont {Minola}, \citenamefont {Keimer}, \citenamefont {Manske},
  \citenamefont {Wang}, \citenamefont {Deinert},\ and\ \citenamefont
  {Kaiser}}]{chu2021fano}%
  \BibitemOpen
  \bibfield  {author} {\bibinfo {author} {\bibfnamefont {H.}~\bibnamefont
  {Chu}}, \bibinfo {author} {\bibfnamefont {S.}~\bibnamefont {Kovalev}},
  \bibinfo {author} {\bibfnamefont {Z.~X.}\ \bibnamefont {Wang}}, \bibinfo
  {author} {\bibfnamefont {L.}~\bibnamefont {Schwarz}}, \bibinfo {author}
  {\bibfnamefont {T.}~\bibnamefont {Dong}}, \bibinfo {author} {\bibfnamefont
  {L.}~\bibnamefont {Feng}}, \bibinfo {author} {\bibfnamefont {R.}~\bibnamefont
  {Haenel}}, \bibinfo {author} {\bibfnamefont {M.-J.}\ \bibnamefont {Kim}},
  \bibinfo {author} {\bibfnamefont {P.}~\bibnamefont {Shabestari}}, \bibinfo
  {author} {\bibfnamefont {H.~L.}\ \bibnamefont {Phuong}}, \bibinfo {author}
  {\bibfnamefont {K.}~\bibnamefont {Honasoge}}, \bibinfo {author}
  {\bibfnamefont {R.~D.}\ \bibnamefont {Dawson}}, \bibinfo {author}
  {\bibfnamefont {D.}~\bibnamefont {Putzky}}, \bibinfo {author} {\bibfnamefont
  {G.}~\bibnamefont {Kim}}, \bibinfo {author} {\bibfnamefont {M.}~\bibnamefont
  {Puviani}}, \bibinfo {author} {\bibfnamefont {M.}~\bibnamefont {Chen}},
  \bibinfo {author} {\bibfnamefont {N.}~\bibnamefont {Awari}}, \bibinfo
  {author} {\bibfnamefont {A.~N.}\ \bibnamefont {Ponomaryov}}, \bibinfo
  {author} {\bibfnamefont {I.}~\bibnamefont {Ilyakov}}, \bibinfo {author}
  {\bibfnamefont {M.}~\bibnamefont {Bluschke}}, \bibinfo {author}
  {\bibfnamefont {F.}~\bibnamefont {Boschini}}, \bibinfo {author}
  {\bibfnamefont {M.}~\bibnamefont {Zonno}}, \bibinfo {author} {\bibfnamefont
  {S.}~\bibnamefont {Zhdanovich}}, \bibinfo {author} {\bibfnamefont
  {M.}~\bibnamefont {Na}}, \bibinfo {author} {\bibfnamefont {G.}~\bibnamefont
  {Christiani}}, \bibinfo {author} {\bibfnamefont {G.}~\bibnamefont
  {Logvenov}}, \bibinfo {author} {\bibfnamefont {D.~J.}\ \bibnamefont {Jones}},
  \bibinfo {author} {\bibfnamefont {A.}~\bibnamefont {Damascelli}}, \bibinfo
  {author} {\bibfnamefont {M.}~\bibnamefont {Minola}}, \bibinfo {author}
  {\bibfnamefont {B.}~\bibnamefont {Keimer}}, \bibinfo {author} {\bibfnamefont
  {D.}~\bibnamefont {Manske}}, \bibinfo {author} {\bibfnamefont
  {N.}~\bibnamefont {Wang}}, \bibinfo {author} {\bibfnamefont {J.-C.}\
  \bibnamefont {Deinert}},\ and\ \bibinfo {author} {\bibfnamefont
  {S.}~\bibnamefont {Kaiser}},\ }\href@noop {} {\bibinfo {title} {Fano
  interference of the higgs mode in cuprate high-tc superconductors}} (\bibinfo
  {year} {2021}),\ \Eprint {https://arxiv.org/abs/2109.09971} {arXiv:2109.09971
  [cond-mat.supr-con]} \BibitemShut {NoStop}%
\bibitem [{\citenamefont {Glier}\ \emph {et~al.}(2023)\citenamefont {Glier},
  \citenamefont {Rerrer}, \citenamefont {Westphal}, \citenamefont {Lüllau},
  \citenamefont {Feng}, \citenamefont {Tian}, \citenamefont {Haenel},
  \citenamefont {Zonno}, \citenamefont {Eisaki}, \citenamefont {Greven},
  \citenamefont {Damascelli}, \citenamefont {Kaiser}, \citenamefont {Manske},\
  and\ \citenamefont {Rübhausen}}]{glier2023direct}%
  \BibitemOpen
  \bibfield  {author} {\bibinfo {author} {\bibfnamefont {T.~E.}\ \bibnamefont
  {Glier}}, \bibinfo {author} {\bibfnamefont {M.}~\bibnamefont {Rerrer}},
  \bibinfo {author} {\bibfnamefont {L.}~\bibnamefont {Westphal}}, \bibinfo
  {author} {\bibfnamefont {G.}~\bibnamefont {Lüllau}}, \bibinfo {author}
  {\bibfnamefont {L.}~\bibnamefont {Feng}}, \bibinfo {author} {\bibfnamefont
  {S.}~\bibnamefont {Tian}}, \bibinfo {author} {\bibfnamefont {R.}~\bibnamefont
  {Haenel}}, \bibinfo {author} {\bibfnamefont {M.}~\bibnamefont {Zonno}},
  \bibinfo {author} {\bibfnamefont {H.}~\bibnamefont {Eisaki}}, \bibinfo
  {author} {\bibfnamefont {M.}~\bibnamefont {Greven}}, \bibinfo {author}
  {\bibfnamefont {A.}~\bibnamefont {Damascelli}}, \bibinfo {author}
  {\bibfnamefont {S.}~\bibnamefont {Kaiser}}, \bibinfo {author} {\bibfnamefont
  {D.}~\bibnamefont {Manske}},\ and\ \bibinfo {author} {\bibfnamefont
  {M.}~\bibnamefont {Rübhausen}},\ }\href@noop {} {\bibinfo {title} {Direct
  observation of the higgs mode in a superconductor by non-equilibrium raman
  scattering}} (\bibinfo {year} {2023}),\ \Eprint
  {https://arxiv.org/abs/2310.08162} {arXiv:2310.08162 [cond-mat.supr-con]}
  \BibitemShut {NoStop}%
\bibitem [{\citenamefont {Cheng}\ \emph {et~al.}(2023)\citenamefont {Cheng},
  \citenamefont {Cheng}, \citenamefont {Lee}, \citenamefont {Mootz},
  \citenamefont {Huang}, \citenamefont {Luo}, \citenamefont {Chen},
  \citenamefont {Lee}, \citenamefont {Wang}, \citenamefont {Perakis},
  \citenamefont {Shen}, \citenamefont {Hwang},\ and\ \citenamefont
  {Wang}}]{cheng2023evidence}%
  \BibitemOpen
  \bibfield  {author} {\bibinfo {author} {\bibfnamefont {B.}~\bibnamefont
  {Cheng}}, \bibinfo {author} {\bibfnamefont {D.}~\bibnamefont {Cheng}},
  \bibinfo {author} {\bibfnamefont {K.}~\bibnamefont {Lee}}, \bibinfo {author}
  {\bibfnamefont {M.}~\bibnamefont {Mootz}}, \bibinfo {author} {\bibfnamefont
  {C.}~\bibnamefont {Huang}}, \bibinfo {author} {\bibfnamefont
  {L.}~\bibnamefont {Luo}}, \bibinfo {author} {\bibfnamefont {.~Z.}\
  \bibnamefont {Chen}}, \bibinfo {author} {\bibfnamefont {Y.}~\bibnamefont
  {Lee}}, \bibinfo {author} {\bibfnamefont {B.~Y.}\ \bibnamefont {Wang}},
  \bibinfo {author} {\bibfnamefont {I.~E.}\ \bibnamefont {Perakis}}, \bibinfo
  {author} {\bibfnamefont {Z.-X.}\ \bibnamefont {Shen}}, \bibinfo {author}
  {\bibfnamefont {H.~Y.}\ \bibnamefont {Hwang}},\ and\ \bibinfo {author}
  {\bibfnamefont {J.}~\bibnamefont {Wang}},\ }\href@noop {} {\bibinfo {title}
  {Evidence for highly damped higgs mode in infinite-layer nickelates}}
  (\bibinfo {year} {2023}),\ \Eprint {https://arxiv.org/abs/2310.02589}
  {arXiv:2310.02589 [cond-mat.supr-con]} \BibitemShut {NoStop}%
\bibitem [{\citenamefont {Harris}\ \emph {et~al.}(1996)\citenamefont {Harris},
  \citenamefont {Shen}, \citenamefont {White}, \citenamefont {Marshall},
  \citenamefont {Schabel}, \citenamefont {Eckstein},\ and\ \citenamefont
  {Bozovic}}]{Harris1996}%
  \BibitemOpen
  \bibfield  {author} {\bibinfo {author} {\bibfnamefont {J.~M.}\ \bibnamefont
  {Harris}}, \bibinfo {author} {\bibfnamefont {Z.~X.}\ \bibnamefont {Shen}},
  \bibinfo {author} {\bibfnamefont {P.~J.}\ \bibnamefont {White}}, \bibinfo
  {author} {\bibfnamefont {D.~S.}\ \bibnamefont {Marshall}}, \bibinfo {author}
  {\bibfnamefont {M.~C.}\ \bibnamefont {Schabel}}, \bibinfo {author}
  {\bibfnamefont {J.~N.}\ \bibnamefont {Eckstein}},\ and\ \bibinfo {author}
  {\bibfnamefont {I.}~\bibnamefont {Bozovic}},\ }\bibfield  {title} {\bibinfo
  {title} {Anomalous superconducting state gap size versus ${T}_{c}$ behavior
  in underdoped
  ${\mathrm{bi}}_{2}{\mathrm{sr}}_{2}{\mathrm{ca}}_{1\ensuremath{-}x}{\mathrm{dy}}_{x}{\mathrm{cu}}_{2}{\mathrm{o}}_{8+\ensuremath{\delta}}$},\
  }\href {https://doi.org/10.1103/PhysRevB.54.R15665} {\bibfield  {journal}
  {\bibinfo  {journal} {Phys. Rev. B}\ }\textbf {\bibinfo {volume} {54}},\
  \bibinfo {pages} {R15665} (\bibinfo {year} {1996})}\BibitemShut {NoStop}%
\bibitem [{\citenamefont {Wang}\ \emph {et~al.}(2005)\citenamefont {Wang},
  \citenamefont {Li}, \citenamefont {Naughton}, \citenamefont {Gu},
  \citenamefont {Uchida},\ and\ \citenamefont {Ong}}]{Wang2005field}%
  \BibitemOpen
  \bibfield  {author} {\bibinfo {author} {\bibfnamefont {Y.}~\bibnamefont
  {Wang}}, \bibinfo {author} {\bibfnamefont {L.}~\bibnamefont {Li}}, \bibinfo
  {author} {\bibfnamefont {M.~J.}\ \bibnamefont {Naughton}}, \bibinfo {author}
  {\bibfnamefont {G.~D.}\ \bibnamefont {Gu}}, \bibinfo {author} {\bibfnamefont
  {S.}~\bibnamefont {Uchida}},\ and\ \bibinfo {author} {\bibfnamefont {N.~P.}\
  \bibnamefont {Ong}},\ }\bibfield  {title} {\bibinfo {title} {Field-enhanced
  diamagnetism in the pseudogap state of the cuprate
  ${\mathrm{bi}}_{2}{\mathrm{sr}}_{2}\mathrm{Ca}{\mathrm{cu}}_{2}{\mathrm{o}}_{8+\ensuremath{\delta}}$
  superconductor in an intense magnetic field},\ }\href
  {https://doi.org/10.1103/PhysRevLett.95.247002} {\bibfield  {journal}
  {\bibinfo  {journal} {Phys. Rev. Lett.}\ }\textbf {\bibinfo {volume} {95}},\
  \bibinfo {pages} {247002} (\bibinfo {year} {2005})}\BibitemShut {NoStop}%
\bibitem [{\citenamefont {Rourke}\ \emph {et~al.}(2011)\citenamefont {Rourke},
  \citenamefont {Mouzopoulou}, \citenamefont {Xu}, \citenamefont
  {Panagopoulos}, \citenamefont {Wang}, \citenamefont {Vignolle}, \citenamefont
  {Proust}, \citenamefont {Kurganova}, \citenamefont {Zeitler}, \citenamefont
  {Tanabe}, \citenamefont {Adachi}, \citenamefont {Koike},\ and\ \citenamefont
  {Hussey}}]{Rourke2011}%
  \BibitemOpen
  \bibfield  {author} {\bibinfo {author} {\bibfnamefont {P.~M.~C.}\
  \bibnamefont {Rourke}}, \bibinfo {author} {\bibfnamefont {I.}~\bibnamefont
  {Mouzopoulou}}, \bibinfo {author} {\bibfnamefont {X.}~\bibnamefont {Xu}},
  \bibinfo {author} {\bibfnamefont {C.}~\bibnamefont {Panagopoulos}}, \bibinfo
  {author} {\bibfnamefont {Y.}~\bibnamefont {Wang}}, \bibinfo {author}
  {\bibfnamefont {B.}~\bibnamefont {Vignolle}}, \bibinfo {author}
  {\bibfnamefont {C.}~\bibnamefont {Proust}}, \bibinfo {author} {\bibfnamefont
  {E.~V.}\ \bibnamefont {Kurganova}}, \bibinfo {author} {\bibfnamefont
  {U.}~\bibnamefont {Zeitler}}, \bibinfo {author} {\bibfnamefont
  {Y.}~\bibnamefont {Tanabe}}, \bibinfo {author} {\bibfnamefont
  {T.}~\bibnamefont {Adachi}}, \bibinfo {author} {\bibfnamefont
  {Y.}~\bibnamefont {Koike}},\ and\ \bibinfo {author} {\bibfnamefont {N.~E.}\
  \bibnamefont {Hussey}},\ }\bibfield  {title} {\bibinfo {title}
  {Phase-fluctuating superconductivity in overdoped la2-{xSrxCuO}4},\ }\href
  {https://doi.org/10.1038/nphys1945} {\bibfield  {journal} {\bibinfo
  {journal} {Nature Physics}\ }\textbf {\bibinfo {volume} {7}},\ \bibinfo
  {pages} {455} (\bibinfo {year} {2011})}\BibitemShut {NoStop}%
\bibitem [{\citenamefont {Schwarz}\ and\ \citenamefont
  {Manske}(2020)}]{schwarz2020theory}%
  \BibitemOpen
  \bibfield  {author} {\bibinfo {author} {\bibfnamefont {L.}~\bibnamefont
  {Schwarz}}\ and\ \bibinfo {author} {\bibfnamefont {D.}~\bibnamefont
  {Manske}},\ }\bibfield  {title} {\bibinfo {title} {Theory of driven higgs
  oscillations and third-harmonic generation in unconventional
  superconductors},\ }\href {https://doi.org/10.1103/PhysRevB.101.184519}
  {\bibfield  {journal} {\bibinfo  {journal} {Phys. Rev. B}\ }\textbf {\bibinfo
  {volume} {101}},\ \bibinfo {pages} {184519} (\bibinfo {year}
  {2020})}\BibitemShut {NoStop}%
\bibitem [{\citenamefont {Cea}\ \emph {et~al.}(2018)\citenamefont {Cea},
  \citenamefont {Barone}, \citenamefont {Castellani},\ and\ \citenamefont
  {Benfatto}}]{Cea2018polarization}%
  \BibitemOpen
  \bibfield  {author} {\bibinfo {author} {\bibfnamefont {T.}~\bibnamefont
  {Cea}}, \bibinfo {author} {\bibfnamefont {P.}~\bibnamefont {Barone}},
  \bibinfo {author} {\bibfnamefont {C.}~\bibnamefont {Castellani}},\ and\
  \bibinfo {author} {\bibfnamefont {L.}~\bibnamefont {Benfatto}},\ }\bibfield
  {title} {\bibinfo {title} {Polarization dependence of the third-harmonic
  generation in multiband superconductors},\ }\href
  {https://doi.org/10.1103/PhysRevB.97.094516} {\bibfield  {journal} {\bibinfo
  {journal} {Phys. Rev. B}\ }\textbf {\bibinfo {volume} {97}},\ \bibinfo
  {pages} {094516} (\bibinfo {year} {2018})}\BibitemShut {NoStop}%
\bibitem [{\citenamefont {Schwarz}\ \emph {et~al.}(2020)\citenamefont
  {Schwarz}, \citenamefont {Fauseweh}, \citenamefont {Tsuji}, \citenamefont
  {Cheng}, \citenamefont {Bittner}, \citenamefont {Krull}, \citenamefont
  {Berciu}, \citenamefont {Uhrig}, \citenamefont {Schnyder}, \citenamefont
  {Kaiser},\ and\ \citenamefont {Manske}}]{Schwarz2020}%
  \BibitemOpen
  \bibfield  {author} {\bibinfo {author} {\bibfnamefont {L.}~\bibnamefont
  {Schwarz}}, \bibinfo {author} {\bibfnamefont {B.}~\bibnamefont {Fauseweh}},
  \bibinfo {author} {\bibfnamefont {N.}~\bibnamefont {Tsuji}}, \bibinfo
  {author} {\bibfnamefont {N.}~\bibnamefont {Cheng}}, \bibinfo {author}
  {\bibfnamefont {N.}~\bibnamefont {Bittner}}, \bibinfo {author} {\bibfnamefont
  {H.}~\bibnamefont {Krull}}, \bibinfo {author} {\bibfnamefont
  {M.}~\bibnamefont {Berciu}}, \bibinfo {author} {\bibfnamefont {G.~S.}\
  \bibnamefont {Uhrig}}, \bibinfo {author} {\bibfnamefont {A.~P.}\ \bibnamefont
  {Schnyder}}, \bibinfo {author} {\bibfnamefont {S.}~\bibnamefont {Kaiser}},\
  and\ \bibinfo {author} {\bibfnamefont {D.}~\bibnamefont {Manske}},\
  }\bibfield  {title} {\bibinfo {title} {Classification and characterization of
  nonequilibrium higgs modes in unconventional superconductors},\ }\bibfield
  {journal} {\bibinfo  {journal} {Nature Communications}\ }\textbf {\bibinfo
  {volume} {11}},\ \href {https://doi.org/10.1038/s41467-019-13763-5}
  {10.1038/s41467-019-13763-5} (\bibinfo {year} {2020})\BibitemShut {NoStop}%
\bibitem [{\citenamefont {Udina}\ \emph {et~al.}(2022)\citenamefont {Udina},
  \citenamefont {Fiore}, \citenamefont {Cea}, \citenamefont {Castellani},
  \citenamefont {Seibold},\ and\ \citenamefont {Benfatto}}]{Udina2022Faraday}%
  \BibitemOpen
  \bibfield  {author} {\bibinfo {author} {\bibfnamefont {M.}~\bibnamefont
  {Udina}}, \bibinfo {author} {\bibfnamefont {J.}~\bibnamefont {Fiore}},
  \bibinfo {author} {\bibfnamefont {T.}~\bibnamefont {Cea}}, \bibinfo {author}
  {\bibfnamefont {C.}~\bibnamefont {Castellani}}, \bibinfo {author}
  {\bibfnamefont {G.}~\bibnamefont {Seibold}},\ and\ \bibinfo {author}
  {\bibfnamefont {L.}~\bibnamefont {Benfatto}},\ }\bibfield  {title} {\bibinfo
  {title} {Thz non-linear optical response in cuprates: predominance of the bcs
  response over the higgs mode},\ }\href {https://doi.org/10.1039/d2fd00016d}
  {\bibfield  {journal} {\bibinfo  {journal} {Faraday Discussions}\ }\textbf
  {\bibinfo {volume} {237}},\ \bibinfo {pages} {168–185} (\bibinfo {year}
  {2022})}\BibitemShut {NoStop}%
\bibitem [{\citenamefont {Alías-Rodríguez}\ \emph {et~al.}(2022)\citenamefont
  {Alías-Rodríguez}, \citenamefont {Basini}, \citenamefont {Benfatto},
  \citenamefont {Boeije}, \citenamefont {Burghardt}, \citenamefont {Burnett},
  \citenamefont {Chen}, \citenamefont {Collet}, \citenamefont {Cowin},
  \citenamefont {Eremin}, \citenamefont {Fleming}, \citenamefont {Girija},
  \citenamefont {Ishida}, \citenamefont {Iwai}, \citenamefont {Johansson},
  \citenamefont {Johnson}, \citenamefont {Katsumi}, \citenamefont {McCusker},
  \citenamefont {Odin}, \citenamefont {Puviani}, \citenamefont {Rost},
  \citenamefont {Rostami}, \citenamefont {Udina},\ and\ \citenamefont
  {Weinstein}}]{AlasRodrguez2022}%
  \BibitemOpen
  \bibfield  {author} {\bibinfo {author} {\bibfnamefont {M.}~\bibnamefont
  {Alías-Rodríguez}}, \bibinfo {author} {\bibfnamefont {M.}~\bibnamefont
  {Basini}}, \bibinfo {author} {\bibfnamefont {L.}~\bibnamefont {Benfatto}},
  \bibinfo {author} {\bibfnamefont {Y.}~\bibnamefont {Boeije}}, \bibinfo
  {author} {\bibfnamefont {I.}~\bibnamefont {Burghardt}}, \bibinfo {author}
  {\bibfnamefont {A.}~\bibnamefont {Burnett}}, \bibinfo {author} {\bibfnamefont
  {L.}~\bibnamefont {Chen}}, \bibinfo {author} {\bibfnamefont {E.}~\bibnamefont
  {Collet}}, \bibinfo {author} {\bibfnamefont {R.}~\bibnamefont {Cowin}},
  \bibinfo {author} {\bibfnamefont {I.}~\bibnamefont {Eremin}}, \bibinfo
  {author} {\bibfnamefont {G.}~\bibnamefont {Fleming}}, \bibinfo {author}
  {\bibfnamefont {A.~V.}\ \bibnamefont {Girija}}, \bibinfo {author}
  {\bibfnamefont {K.}~\bibnamefont {Ishida}}, \bibinfo {author} {\bibfnamefont
  {S.}~\bibnamefont {Iwai}}, \bibinfo {author} {\bibfnamefont {J.~O.}\
  \bibnamefont {Johansson}}, \bibinfo {author} {\bibfnamefont {S.~L.}\
  \bibnamefont {Johnson}}, \bibinfo {author} {\bibfnamefont {K.}~\bibnamefont
  {Katsumi}}, \bibinfo {author} {\bibfnamefont {J.}~\bibnamefont {McCusker}},
  \bibinfo {author} {\bibfnamefont {C.}~\bibnamefont {Odin}}, \bibinfo {author}
  {\bibfnamefont {M.}~\bibnamefont {Puviani}}, \bibinfo {author} {\bibfnamefont
  {J.~M.}\ \bibnamefont {Rost}}, \bibinfo {author} {\bibfnamefont
  {H.}~\bibnamefont {Rostami}}, \bibinfo {author} {\bibfnamefont
  {M.}~\bibnamefont {Udina}},\ and\ \bibinfo {author} {\bibfnamefont
  {J.}~\bibnamefont {Weinstein}},\ }\bibfield  {title} {\bibinfo {title}
  {Theory of out of equilibrium light-induced phenomena: general discussion},\
  }\href {https://doi.org/10.1039/d2fd90036j} {\bibfield  {journal} {\bibinfo
  {journal} {Faraday Discussions}\ }\textbf {\bibinfo {volume} {237}},\
  \bibinfo {pages} {198–223} (\bibinfo {year} {2022})}\BibitemShut {NoStop}%
\bibitem [{\citenamefont {Katsumi}\ \emph {et~al.}(2023)\citenamefont
  {Katsumi}, \citenamefont {Fiore}, \citenamefont {Udina}, \citenamefont {au2},
  \citenamefont {Barbalas}, \citenamefont {Jesudasan}, \citenamefont
  {Raychaudhuri}, \citenamefont {Seibold}, \citenamefont {Benfatto},\ and\
  \citenamefont {Armitage}}]{katsumi2023revealing}%
  \BibitemOpen
  \bibfield  {author} {\bibinfo {author} {\bibfnamefont {K.}~\bibnamefont
  {Katsumi}}, \bibinfo {author} {\bibfnamefont {J.}~\bibnamefont {Fiore}},
  \bibinfo {author} {\bibfnamefont {M.}~\bibnamefont {Udina}}, \bibinfo
  {author} {\bibfnamefont {R.~R.~I.}\ \bibnamefont {au2}}, \bibinfo {author}
  {\bibfnamefont {D.}~\bibnamefont {Barbalas}}, \bibinfo {author}
  {\bibfnamefont {J.}~\bibnamefont {Jesudasan}}, \bibinfo {author}
  {\bibfnamefont {P.}~\bibnamefont {Raychaudhuri}}, \bibinfo {author}
  {\bibfnamefont {G.}~\bibnamefont {Seibold}}, \bibinfo {author} {\bibfnamefont
  {L.}~\bibnamefont {Benfatto}},\ and\ \bibinfo {author} {\bibfnamefont
  {N.~P.}\ \bibnamefont {Armitage}},\ }\href@noop {} {\bibinfo {title}
  {Revealing novel aspects of light-matter coupling in terahertz
  two-dimensional coherent spectroscopy: the case of the amplitude mode in
  superconductors}} (\bibinfo {year} {2023}),\ \Eprint
  {https://arxiv.org/abs/2311.16449} {arXiv:2311.16449 [cond-mat.supr-con]}
  \BibitemShut {NoStop}%
\bibitem [{\citenamefont {Giusti}\ \emph {et~al.}(2019)\citenamefont {Giusti},
  \citenamefont {Marciniak}, \citenamefont {Randi}, \citenamefont {Sparapassi},
  \citenamefont {Boschini}, \citenamefont {Eisaki}, \citenamefont {Greven},
  \citenamefont {Damascelli}, \citenamefont {Avella},\ and\ \citenamefont
  {Fausti}}]{Giusti2019signatures}%
  \BibitemOpen
  \bibfield  {author} {\bibinfo {author} {\bibfnamefont {F.}~\bibnamefont
  {Giusti}}, \bibinfo {author} {\bibfnamefont {A.}~\bibnamefont {Marciniak}},
  \bibinfo {author} {\bibfnamefont {F.}~\bibnamefont {Randi}}, \bibinfo
  {author} {\bibfnamefont {G.}~\bibnamefont {Sparapassi}}, \bibinfo {author}
  {\bibfnamefont {F.}~\bibnamefont {Boschini}}, \bibinfo {author}
  {\bibfnamefont {H.}~\bibnamefont {Eisaki}}, \bibinfo {author} {\bibfnamefont
  {M.}~\bibnamefont {Greven}}, \bibinfo {author} {\bibfnamefont
  {A.}~\bibnamefont {Damascelli}}, \bibinfo {author} {\bibfnamefont
  {A.}~\bibnamefont {Avella}},\ and\ \bibinfo {author} {\bibfnamefont
  {D.}~\bibnamefont {Fausti}},\ }\bibfield  {title} {\bibinfo {title}
  {Signatures of enhanced superconducting phase coherence in optimally doped
  ${\mathrm{bi}}_{2}{\mathrm{sr}}_{2}{\mathrm{y}}_{0.08}{\mathrm{ca}}_{0.92}{\mathrm{cu}}_{2}{\mathrm{o}}_{8+\ensuremath{\delta}}$
  driven by midinfrared pulse excitations},\ }\href
  {https://doi.org/10.1103/PhysRevLett.122.067002} {\bibfield  {journal}
  {\bibinfo  {journal} {Phys. Rev. Lett.}\ }\textbf {\bibinfo {volume} {122}},\
  \bibinfo {pages} {067002} (\bibinfo {year} {2019})}\BibitemShut {NoStop}%
\bibitem [{\citenamefont {Emery}\ and\ \citenamefont
  {Kivelson}(1995)}]{Emery1995}%
  \BibitemOpen
  \bibfield  {author} {\bibinfo {author} {\bibfnamefont {V.~J.}\ \bibnamefont
  {Emery}}\ and\ \bibinfo {author} {\bibfnamefont {S.~A.}\ \bibnamefont
  {Kivelson}},\ }\bibfield  {title} {\bibinfo {title} {Importance of phase
  fluctuations in superconductors with small superfluid density},\ }\href
  {https://doi.org/10.1038/374434a0} {\bibfield  {journal} {\bibinfo  {journal}
  {Nature}\ }\textbf {\bibinfo {volume} {374}},\ \bibinfo {pages} {434–437}
  (\bibinfo {year} {1995})}\BibitemShut {NoStop}%
\bibitem [{\citenamefont {Corson}\ \emph {et~al.}(1999)\citenamefont {Corson},
  \citenamefont {Mallozzi}, \citenamefont {Orenstein}, \citenamefont
  {Eckstein},\ and\ \citenamefont {Bozovic}}]{Corson1999}%
  \BibitemOpen
  \bibfield  {author} {\bibinfo {author} {\bibfnamefont {J.}~\bibnamefont
  {Corson}}, \bibinfo {author} {\bibfnamefont {R.}~\bibnamefont {Mallozzi}},
  \bibinfo {author} {\bibfnamefont {J.}~\bibnamefont {Orenstein}}, \bibinfo
  {author} {\bibfnamefont {J.~N.}\ \bibnamefont {Eckstein}},\ and\ \bibinfo
  {author} {\bibfnamefont {I.}~\bibnamefont {Bozovic}},\ }\bibfield  {title}
  {\bibinfo {title} {Vanishing of phase coherence in underdoped
  bi2sr2cacu2o8},\ }\href {https://doi.org/10.1038/18402} {\bibfield
  {journal} {\bibinfo  {journal} {Nature}\ }\textbf {\bibinfo {volume} {398}},\
  \bibinfo {pages} {221–223} (\bibinfo {year} {1999})}\BibitemShut {NoStop}%
\bibitem [{\citenamefont {Xu}\ \emph {et~al.}(2000)\citenamefont {Xu},
  \citenamefont {Ong}, \citenamefont {Wang}, \citenamefont {Kakeshita},\ and\
  \citenamefont {Uchida}}]{Xu2000}%
  \BibitemOpen
  \bibfield  {author} {\bibinfo {author} {\bibfnamefont {Z.~A.}\ \bibnamefont
  {Xu}}, \bibinfo {author} {\bibfnamefont {N.~P.}\ \bibnamefont {Ong}},
  \bibinfo {author} {\bibfnamefont {Y.}~\bibnamefont {Wang}}, \bibinfo {author}
  {\bibfnamefont {T.}~\bibnamefont {Kakeshita}},\ and\ \bibinfo {author}
  {\bibfnamefont {S.}~\bibnamefont {Uchida}},\ }\bibfield  {title} {\bibinfo
  {title} {Vortex-like excitations and the onset of superconducting phase
  fluctuation in underdoped la2-xsrxcuo4},\ }\href
  {https://doi.org/10.1038/35020016} {\bibfield  {journal} {\bibinfo  {journal}
  {Nature}\ }\textbf {\bibinfo {volume} {406}},\ \bibinfo {pages} {486–488}
  (\bibinfo {year} {2000})}\BibitemShut {NoStop}%
\bibitem [{\citenamefont {Giusti}\ \emph {et~al.}(2021)\citenamefont {Giusti},
  \citenamefont {Montanaro}, \citenamefont {Marciniak}, \citenamefont {Randi},
  \citenamefont {Boschini}, \citenamefont {Glerean}, \citenamefont {Jarc},
  \citenamefont {Eisaki}, \citenamefont {Greven}, \citenamefont {Damascelli},
  \citenamefont {Avella},\ and\ \citenamefont
  {Fausti}}]{Giusti2021anisotropic}%
  \BibitemOpen
  \bibfield  {author} {\bibinfo {author} {\bibfnamefont {F.}~\bibnamefont
  {Giusti}}, \bibinfo {author} {\bibfnamefont {A.}~\bibnamefont {Montanaro}},
  \bibinfo {author} {\bibfnamefont {A.}~\bibnamefont {Marciniak}}, \bibinfo
  {author} {\bibfnamefont {F.}~\bibnamefont {Randi}}, \bibinfo {author}
  {\bibfnamefont {F.}~\bibnamefont {Boschini}}, \bibinfo {author}
  {\bibfnamefont {F.}~\bibnamefont {Glerean}}, \bibinfo {author} {\bibfnamefont
  {G.}~\bibnamefont {Jarc}}, \bibinfo {author} {\bibfnamefont {H.}~\bibnamefont
  {Eisaki}}, \bibinfo {author} {\bibfnamefont {M.}~\bibnamefont {Greven}},
  \bibinfo {author} {\bibfnamefont {A.}~\bibnamefont {Damascelli}}, \bibinfo
  {author} {\bibfnamefont {A.}~\bibnamefont {Avella}},\ and\ \bibinfo {author}
  {\bibfnamefont {D.}~\bibnamefont {Fausti}},\ }\bibfield  {title} {\bibinfo
  {title} {Anisotropic time-domain electronic response in cuprates driven by
  midinfrared pulses},\ }\href {https://doi.org/10.1103/PhysRevB.104.125121}
  {\bibfield  {journal} {\bibinfo  {journal} {Phys. Rev. B}\ }\textbf {\bibinfo
  {volume} {104}},\ \bibinfo {pages} {125121} (\bibinfo {year}
  {2021})}\BibitemShut {NoStop}%
\bibitem [{\citenamefont {Buzzi}\ \emph {et~al.}(2021)\citenamefont {Buzzi},
  \citenamefont {Jotzu}, \citenamefont {Cavalleri}, \citenamefont {Cirac},
  \citenamefont {Demler}, \citenamefont {Halperin}, \citenamefont {Lukin},
  \citenamefont {Shi}, \citenamefont {Wang},\ and\ \citenamefont
  {Podolsky}}]{Buzzi2021Higgs}%
  \BibitemOpen
  \bibfield  {author} {\bibinfo {author} {\bibfnamefont {M.}~\bibnamefont
  {Buzzi}}, \bibinfo {author} {\bibfnamefont {G.}~\bibnamefont {Jotzu}},
  \bibinfo {author} {\bibfnamefont {A.}~\bibnamefont {Cavalleri}}, \bibinfo
  {author} {\bibfnamefont {J.~I.}\ \bibnamefont {Cirac}}, \bibinfo {author}
  {\bibfnamefont {E.~A.}\ \bibnamefont {Demler}}, \bibinfo {author}
  {\bibfnamefont {B.~I.}\ \bibnamefont {Halperin}}, \bibinfo {author}
  {\bibfnamefont {M.~D.}\ \bibnamefont {Lukin}}, \bibinfo {author}
  {\bibfnamefont {T.}~\bibnamefont {Shi}}, \bibinfo {author} {\bibfnamefont
  {Y.}~\bibnamefont {Wang}},\ and\ \bibinfo {author} {\bibfnamefont
  {D.}~\bibnamefont {Podolsky}},\ }\bibfield  {title} {\bibinfo {title}
  {Higgs-mediated optical amplification in a nonequilibrium superconductor},\
  }\href {https://doi.org/10.1103/PhysRevX.11.011055} {\bibfield  {journal}
  {\bibinfo  {journal} {Phys. Rev. X}\ }\textbf {\bibinfo {volume} {11}},\
  \bibinfo {pages} {011055} (\bibinfo {year} {2021})}\BibitemShut {NoStop}%
\bibitem [{\citenamefont {Udina}\ \emph {et~al.}(2019)\citenamefont {Udina},
  \citenamefont {Cea},\ and\ \citenamefont {Benfatto}}]{Udina2019theory}%
  \BibitemOpen
  \bibfield  {author} {\bibinfo {author} {\bibfnamefont {M.}~\bibnamefont
  {Udina}}, \bibinfo {author} {\bibfnamefont {T.}~\bibnamefont {Cea}},\ and\
  \bibinfo {author} {\bibfnamefont {L.}~\bibnamefont {Benfatto}},\ }\bibfield
  {title} {\bibinfo {title} {{Theory of coherent-oscillations generation in
  terahertz pump-probe spectroscopy: From phonons to electronic collective
  modes}},\ }\href {https://doi.org/10.1103/PhysRevB.100.165131} {\bibfield
  {journal} {\bibinfo  {journal} {Phys. Rev. B}\ }\textbf {\bibinfo {volume}
  {100}},\ \bibinfo {pages} {165131} (\bibinfo {year} {2019})}\BibitemShut
  {NoStop}%
\bibitem [{\citenamefont {Giorgianni}\ \emph {et~al.}(2019)\citenamefont
  {Giorgianni}, \citenamefont {Cea}, \citenamefont {Vicario}, \citenamefont
  {Hauri}, \citenamefont {Withanage}, \citenamefont {Xi},\ and\ \citenamefont
  {Benfatto}}]{Giorgianni2019}%
  \BibitemOpen
  \bibfield  {author} {\bibinfo {author} {\bibfnamefont {F.}~\bibnamefont
  {Giorgianni}}, \bibinfo {author} {\bibfnamefont {T.}~\bibnamefont {Cea}},
  \bibinfo {author} {\bibfnamefont {C.}~\bibnamefont {Vicario}}, \bibinfo
  {author} {\bibfnamefont {C.~P.}\ \bibnamefont {Hauri}}, \bibinfo {author}
  {\bibfnamefont {W.~K.}\ \bibnamefont {Withanage}}, \bibinfo {author}
  {\bibfnamefont {X.}~\bibnamefont {Xi}},\ and\ \bibinfo {author}
  {\bibfnamefont {L.}~\bibnamefont {Benfatto}},\ }\bibfield  {title} {\bibinfo
  {title} {{Leggett mode controlled by light pulses}},\ }\href
  {https://doi.org/10.1038/s41567-018-0385-4} {\bibfield  {journal} {\bibinfo
  {journal} {Nature Physics}\ }\textbf {\bibinfo {volume} {15}},\ \bibinfo
  {pages} {341–346} (\bibinfo {year} {2019})}\BibitemShut {NoStop}%
\bibitem [{\citenamefont {Cundiff}\ and\ \citenamefont
  {Mukamel}(2013)}]{Cundiff2013}%
  \BibitemOpen
  \bibfield  {author} {\bibinfo {author} {\bibfnamefont {S.~T.}\ \bibnamefont
  {Cundiff}}\ and\ \bibinfo {author} {\bibfnamefont {S.}~\bibnamefont
  {Mukamel}},\ }\bibfield  {title} {\bibinfo {title} {Optical multidimensional
  coherent spectroscopy},\ }\href {https://doi.org/10.1063/pt.3.2047}
  {\bibfield  {journal} {\bibinfo  {journal} {Physics Today}\ }\textbf
  {\bibinfo {volume} {66}},\ \bibinfo {pages} {44} (\bibinfo {year}
  {2013})}\BibitemShut {NoStop}%
\bibitem [{\citenamefont {Woerner}\ \emph {et~al.}(2013)\citenamefont
  {Woerner}, \citenamefont {Kuehn}, \citenamefont {Bowlan}, \citenamefont
  {Reimann},\ and\ \citenamefont {Elsaesser}}]{Woerner2013}%
  \BibitemOpen
  \bibfield  {author} {\bibinfo {author} {\bibfnamefont {M.}~\bibnamefont
  {Woerner}}, \bibinfo {author} {\bibfnamefont {W.}~\bibnamefont {Kuehn}},
  \bibinfo {author} {\bibfnamefont {P.}~\bibnamefont {Bowlan}}, \bibinfo
  {author} {\bibfnamefont {K.}~\bibnamefont {Reimann}},\ and\ \bibinfo {author}
  {\bibfnamefont {T.}~\bibnamefont {Elsaesser}},\ }\bibfield  {title} {\bibinfo
  {title} {Ultrafast two-dimensional terahertz spectroscopy of elementary
  excitations in solids},\ }\href
  {https://doi.org/10.1088/1367-2630/15/2/025039} {\bibfield  {journal}
  {\bibinfo  {journal} {New Journal of Physics}\ }\textbf {\bibinfo {volume}
  {15}},\ \bibinfo {pages} {025039} (\bibinfo {year} {2013})}\BibitemShut
  {NoStop}%
\bibitem [{\citenamefont {Lu}\ \emph {et~al.}(2017)\citenamefont {Lu},
  \citenamefont {Li}, \citenamefont {Hwang}, \citenamefont {Ofori-Okai},
  \citenamefont {Kurihara}, \citenamefont {Suemoto},\ and\ \citenamefont
  {Nelson}}]{PhysRevLett.118.207204}%
  \BibitemOpen
  \bibfield  {author} {\bibinfo {author} {\bibfnamefont {J.}~\bibnamefont
  {Lu}}, \bibinfo {author} {\bibfnamefont {X.}~\bibnamefont {Li}}, \bibinfo
  {author} {\bibfnamefont {H.~Y.}\ \bibnamefont {Hwang}}, \bibinfo {author}
  {\bibfnamefont {B.~K.}\ \bibnamefont {Ofori-Okai}}, \bibinfo {author}
  {\bibfnamefont {T.}~\bibnamefont {Kurihara}}, \bibinfo {author}
  {\bibfnamefont {T.}~\bibnamefont {Suemoto}},\ and\ \bibinfo {author}
  {\bibfnamefont {K.~A.}\ \bibnamefont {Nelson}},\ }\bibfield  {title}
  {\bibinfo {title} {Coherent two-dimensional terahertz magnetic resonance
  spectroscopy of collective spin waves},\ }\href
  {https://doi.org/10.1103/PhysRevLett.118.207204} {\bibfield  {journal}
  {\bibinfo  {journal} {Phys. Rev. Lett.}\ }\textbf {\bibinfo {volume} {118}},\
  \bibinfo {pages} {207204} (\bibinfo {year} {2017})}\BibitemShut {NoStop}%
\bibitem [{\citenamefont {Wan}\ and\ \citenamefont
  {Armitage}(2019)}]{PhysRevLett.122.257401}%
  \BibitemOpen
  \bibfield  {author} {\bibinfo {author} {\bibfnamefont {Y.}~\bibnamefont
  {Wan}}\ and\ \bibinfo {author} {\bibfnamefont {N.~P.}\ \bibnamefont
  {Armitage}},\ }\bibfield  {title} {\bibinfo {title} {Resolving continua of
  fractional excitations by spinon echo in thz 2d coherent spectroscopy},\
  }\href {https://doi.org/10.1103/PhysRevLett.122.257401} {\bibfield  {journal}
  {\bibinfo  {journal} {Phys. Rev. Lett.}\ }\textbf {\bibinfo {volume} {122}},\
  \bibinfo {pages} {257401} (\bibinfo {year} {2019})}\BibitemShut {NoStop}%
\bibitem [{\citenamefont {Mahmood}\ \emph {et~al.}(2021)\citenamefont
  {Mahmood}, \citenamefont {Chaudhuri}, \citenamefont {Gopalakrishnan},
  \citenamefont {Nandkishore},\ and\ \citenamefont {Armitage}}]{Mahmood2021}%
  \BibitemOpen
  \bibfield  {author} {\bibinfo {author} {\bibfnamefont {F.}~\bibnamefont
  {Mahmood}}, \bibinfo {author} {\bibfnamefont {D.}~\bibnamefont {Chaudhuri}},
  \bibinfo {author} {\bibfnamefont {S.}~\bibnamefont {Gopalakrishnan}},
  \bibinfo {author} {\bibfnamefont {R.}~\bibnamefont {Nandkishore}},\ and\
  \bibinfo {author} {\bibfnamefont {N.~P.}\ \bibnamefont {Armitage}},\
  }\bibfield  {title} {\bibinfo {title} {Observation of a marginal fermi
  glass},\ }\href {https://doi.org/10.1038/s41567-020-01149-0} {\bibfield
  {journal} {\bibinfo  {journal} {Nature Physics}\ }\textbf {\bibinfo {volume}
  {17}},\ \bibinfo {pages} {627} (\bibinfo {year} {2021})}\BibitemShut
  {NoStop}%
\bibitem [{\citenamefont {Puviani}\ and\ \citenamefont
  {Manske}(2022)}]{Puviani2022}%
  \BibitemOpen
  \bibfield  {author} {\bibinfo {author} {\bibfnamefont {M.}~\bibnamefont
  {Puviani}}\ and\ \bibinfo {author} {\bibfnamefont {D.}~\bibnamefont
  {Manske}},\ }\bibfield  {title} {\bibinfo {title} {Quench-drive spectroscopy
  of cuprates},\ }\bibfield  {journal} {\bibinfo  {journal} {Faraday
  Discussions}\ }\href {https://doi.org/10.1039/d2fd00010e}
  {10.1039/d2fd00010e} (\bibinfo {year} {2022})\BibitemShut {NoStop}%
\bibitem [{\citenamefont {Puviani}\ \emph {et~al.}(2023)\citenamefont
  {Puviani}, \citenamefont {Haenel},\ and\ \citenamefont
  {Manske}}]{Puviani2023}%
  \BibitemOpen
  \bibfield  {author} {\bibinfo {author} {\bibfnamefont {M.}~\bibnamefont
  {Puviani}}, \bibinfo {author} {\bibfnamefont {R.}~\bibnamefont {Haenel}},\
  and\ \bibinfo {author} {\bibfnamefont {D.}~\bibnamefont {Manske}},\
  }\bibfield  {title} {\bibinfo {title} {Quench-drive spectroscopy and
  high-harmonic generation in bcs superconductors},\ }\href
  {https://doi.org/10.1103/PhysRevB.107.094501} {\bibfield  {journal} {\bibinfo
   {journal} {Phys. Rev. B}\ }\textbf {\bibinfo {volume} {107}},\ \bibinfo
  {pages} {094501} (\bibinfo {year} {2023})}\BibitemShut {NoStop}%
\bibitem [{\citenamefont {Kim}\ \emph {et~al.}(2023)\citenamefont {Kim},
  \citenamefont {Kovalev}, \citenamefont {Udina}, \citenamefont {Haenel},
  \citenamefont {Kim}, \citenamefont {Puviani}, \citenamefont {Cristiani},
  \citenamefont {Ilyakov}, \citenamefont {de~Oliveira}, \citenamefont
  {Ponomaryov}, \citenamefont {Deinert}, \citenamefont {Logvenov},
  \citenamefont {Keimer}, \citenamefont {Manske}, \citenamefont {Benfatto},\
  and\ \citenamefont {Kaiser}}]{kim2023tracing}%
  \BibitemOpen
  \bibfield  {author} {\bibinfo {author} {\bibfnamefont {M.-J.}\ \bibnamefont
  {Kim}}, \bibinfo {author} {\bibfnamefont {S.}~\bibnamefont {Kovalev}},
  \bibinfo {author} {\bibfnamefont {M.}~\bibnamefont {Udina}}, \bibinfo
  {author} {\bibfnamefont {R.}~\bibnamefont {Haenel}}, \bibinfo {author}
  {\bibfnamefont {G.}~\bibnamefont {Kim}}, \bibinfo {author} {\bibfnamefont
  {M.}~\bibnamefont {Puviani}}, \bibinfo {author} {\bibfnamefont
  {G.}~\bibnamefont {Cristiani}}, \bibinfo {author} {\bibfnamefont
  {I.}~\bibnamefont {Ilyakov}}, \bibinfo {author} {\bibfnamefont {T.~V. A.~G.}\
  \bibnamefont {de~Oliveira}}, \bibinfo {author} {\bibfnamefont
  {A.}~\bibnamefont {Ponomaryov}}, \bibinfo {author} {\bibfnamefont {J.-C.}\
  \bibnamefont {Deinert}}, \bibinfo {author} {\bibfnamefont {G.}~\bibnamefont
  {Logvenov}}, \bibinfo {author} {\bibfnamefont {B.}~\bibnamefont {Keimer}},
  \bibinfo {author} {\bibfnamefont {D.}~\bibnamefont {Manske}}, \bibinfo
  {author} {\bibfnamefont {L.}~\bibnamefont {Benfatto}},\ and\ \bibinfo
  {author} {\bibfnamefont {S.}~\bibnamefont {Kaiser}},\ }\href@noop {}
  {\bibinfo {title} {Tracing the dynamics of superconducting order via
  transient third harmonic generation}} (\bibinfo {year} {2023}),\ \Eprint
  {https://arxiv.org/abs/2303.03288} {arXiv:2303.03288 [cond-mat.supr-con]}
  \BibitemShut {NoStop}%
\bibitem [{\citenamefont {Mootz}\ \emph {et~al.}(2022)\citenamefont {Mootz},
  \citenamefont {Luo}, \citenamefont {Wang},\ and\ \citenamefont
  {Perakis}}]{Mootz2022visualization}%
  \BibitemOpen
  \bibfield  {author} {\bibinfo {author} {\bibfnamefont {M.}~\bibnamefont
  {Mootz}}, \bibinfo {author} {\bibfnamefont {L.}~\bibnamefont {Luo}}, \bibinfo
  {author} {\bibfnamefont {J.}~\bibnamefont {Wang}},\ and\ \bibinfo {author}
  {\bibfnamefont {l.~E.}\ \bibnamefont {Perakis}},\ }\bibfield  {title}
  {\bibinfo {title} {Visualization and quantum control of light-accelerated
  condensates by terahertz multi-dimensional coherent spectroscopy},\
  }\bibfield  {journal} {\bibinfo  {journal} {Communications Physics}\ }\textbf
  {\bibinfo {volume} {5}},\ \href {https://doi.org/10.1038/s42005-022-00822-5}
  {10.1038/s42005-022-00822-5} (\bibinfo {year} {2022})\BibitemShut {NoStop}%
\bibitem [{\citenamefont {Luo}\ \emph {et~al.}(2022)\citenamefont {Luo},
  \citenamefont {Mootz}, \citenamefont {Kang}, \citenamefont {Huang},
  \citenamefont {Eom}, \citenamefont {Lee}, \citenamefont {Vaswani},
  \citenamefont {Collantes}, \citenamefont {Hellstrom}, \citenamefont
  {Perakis}, \citenamefont {Eom},\ and\ \citenamefont {Wang}}]{Luo2022}%
  \BibitemOpen
  \bibfield  {author} {\bibinfo {author} {\bibfnamefont {L.}~\bibnamefont
  {Luo}}, \bibinfo {author} {\bibfnamefont {M.}~\bibnamefont {Mootz}}, \bibinfo
  {author} {\bibfnamefont {J.~H.}\ \bibnamefont {Kang}}, \bibinfo {author}
  {\bibfnamefont {C.}~\bibnamefont {Huang}}, \bibinfo {author} {\bibfnamefont
  {K.}~\bibnamefont {Eom}}, \bibinfo {author} {\bibfnamefont {J.~W.}\
  \bibnamefont {Lee}}, \bibinfo {author} {\bibfnamefont {C.}~\bibnamefont
  {Vaswani}}, \bibinfo {author} {\bibfnamefont {Y.~G.}\ \bibnamefont
  {Collantes}}, \bibinfo {author} {\bibfnamefont {E.~E.}\ \bibnamefont
  {Hellstrom}}, \bibinfo {author} {\bibfnamefont {I.~E.}\ \bibnamefont
  {Perakis}}, \bibinfo {author} {\bibfnamefont {C.~B.}\ \bibnamefont {Eom}},\
  and\ \bibinfo {author} {\bibfnamefont {J.}~\bibnamefont {Wang}},\ }\bibfield
  {title} {\bibinfo {title} {Quantum coherence tomography of light-controlled
  superconductivity},\ }\bibfield  {journal} {\bibinfo  {journal} {Nature
  Physics}\ }\href {https://doi.org/10.1038/s41567-022-01827-1}
  {10.1038/s41567-022-01827-1} (\bibinfo {year} {2022})\BibitemShut {NoStop}%
\bibitem [{\citenamefont {Mootz}\ \emph {et~al.}(2023)\citenamefont {Mootz},
  \citenamefont {Orth}, \citenamefont {Huang}, \citenamefont {Luo},
  \citenamefont {Wang},\ and\ \citenamefont {Yao}}]{mootz2023twodimensional}%
  \BibitemOpen
  \bibfield  {author} {\bibinfo {author} {\bibfnamefont {M.}~\bibnamefont
  {Mootz}}, \bibinfo {author} {\bibfnamefont {P.~P.}\ \bibnamefont {Orth}},
  \bibinfo {author} {\bibfnamefont {C.}~\bibnamefont {Huang}}, \bibinfo
  {author} {\bibfnamefont {L.}~\bibnamefont {Luo}}, \bibinfo {author}
  {\bibfnamefont {J.}~\bibnamefont {Wang}},\ and\ \bibinfo {author}
  {\bibfnamefont {Y.-X.}\ \bibnamefont {Yao}},\ }\href@noop {} {\bibinfo
  {title} {Two-dimensional coherent spectrum of high-spin models via a quantum
  computing approach}} (\bibinfo {year} {2023}),\ \Eprint
  {https://arxiv.org/abs/2311.14035} {arXiv:2311.14035 [quant-ph]} \BibitemShut
  {NoStop}%
\bibitem [{\citenamefont {Mootz}\ \emph {et~al.}(2024)\citenamefont {Mootz},
  \citenamefont {Luo}, \citenamefont {Huang}, \citenamefont {Wang},\ and\
  \citenamefont {llias E.~Perakis}}]{mootz2024multidimensional}%
  \BibitemOpen
  \bibfield  {author} {\bibinfo {author} {\bibfnamefont {M.}~\bibnamefont
  {Mootz}}, \bibinfo {author} {\bibfnamefont {L.}~\bibnamefont {Luo}}, \bibinfo
  {author} {\bibfnamefont {C.}~\bibnamefont {Huang}}, \bibinfo {author}
  {\bibfnamefont {J.}~\bibnamefont {Wang}},\ and\ \bibinfo {author}
  {\bibnamefont {llias E.~Perakis}},\ }\href@noop {} {\bibinfo {title}
  {Multi-dimensional coherent spectroscopy of light-driven states and their
  collective modes in multi-band superconductors}} (\bibinfo {year} {2024}),\
  \Eprint {https://arxiv.org/abs/2310.03950} {arXiv:2310.03950
  [cond-mat.supr-con]} \BibitemShut {NoStop}%
\bibitem [{\citenamefont {Salvador}\ \emph {et~al.}(2024)\citenamefont
  {Salvador}, \citenamefont {Dolgirev}, \citenamefont {Michael}, \citenamefont
  {Liu}, \citenamefont {Pavicevic}, \citenamefont {Fechner}, \citenamefont
  {Cavalleri},\ and\ \citenamefont {Demler}}]{salvador2024principles}%
  \BibitemOpen
  \bibfield  {author} {\bibinfo {author} {\bibfnamefont {A.~G.}\ \bibnamefont
  {Salvador}}, \bibinfo {author} {\bibfnamefont {P.~E.}\ \bibnamefont
  {Dolgirev}}, \bibinfo {author} {\bibfnamefont {M.~H.}\ \bibnamefont
  {Michael}}, \bibinfo {author} {\bibfnamefont {A.}~\bibnamefont {Liu}},
  \bibinfo {author} {\bibfnamefont {D.}~\bibnamefont {Pavicevic}}, \bibinfo
  {author} {\bibfnamefont {M.}~\bibnamefont {Fechner}}, \bibinfo {author}
  {\bibfnamefont {A.}~\bibnamefont {Cavalleri}},\ and\ \bibinfo {author}
  {\bibfnamefont {E.}~\bibnamefont {Demler}},\ }\href@noop {} {\bibinfo {title}
  {Principles of 2d terahertz spectroscopy of collective excitations: the case
  of josephson plasmons in layered superconductors}} (\bibinfo {year} {2024}),\
  \Eprint {https://arxiv.org/abs/2401.05503} {arXiv:2401.05503
  [cond-mat.supr-con]} \BibitemShut {NoStop}%
\bibitem [{\citenamefont {Devereaux}\ and\ \citenamefont
  {Einzel}(1995)}]{Devereaux1995}%
  \BibitemOpen
  \bibfield  {author} {\bibinfo {author} {\bibfnamefont {T.~P.}\ \bibnamefont
  {Devereaux}}\ and\ \bibinfo {author} {\bibfnamefont {D.}~\bibnamefont
  {Einzel}},\ }\bibfield  {title} {\bibinfo {title} {Electronic raman
  scattering in superconductors as a probe of anisotropic electron pairing},\
  }\href {https://doi.org/10.1103/PhysRevB.51.16336} {\bibfield  {journal}
  {\bibinfo  {journal} {Phys. Rev. B}\ }\textbf {\bibinfo {volume} {51}},\
  \bibinfo {pages} {16336} (\bibinfo {year} {1995})}\BibitemShut {NoStop}%
\bibitem [{\citenamefont {Devereaux}\ and\ \citenamefont
  {Hackl}(2007)}]{Devereaux2007inelastic}%
  \BibitemOpen
  \bibfield  {author} {\bibinfo {author} {\bibfnamefont {T.~P.}\ \bibnamefont
  {Devereaux}}\ and\ \bibinfo {author} {\bibfnamefont {R.}~\bibnamefont
  {Hackl}},\ }\bibfield  {title} {\bibinfo {title} {Inelastic light scattering
  from correlated electrons},\ }\href
  {https://doi.org/10.1103/RevModPhys.79.175} {\bibfield  {journal} {\bibinfo
  {journal} {Rev. Mod. Phys.}\ }\textbf {\bibinfo {volume} {79}},\ \bibinfo
  {pages} {175} (\bibinfo {year} {2007})}\BibitemShut {NoStop}%
\bibitem [{\citenamefont {Kondo}\ \emph {et~al.}(2015)\citenamefont {Kondo},
  \citenamefont {Malaeb}, \citenamefont {Ishida}, \citenamefont {Sasagawa},
  \citenamefont {Sakamoto}, \citenamefont {Takeuchi}, \citenamefont {Tohyama},\
  and\ \citenamefont {Shin}}]{Kondo2015}%
  \BibitemOpen
  \bibfield  {author} {\bibinfo {author} {\bibfnamefont {T.}~\bibnamefont
  {Kondo}}, \bibinfo {author} {\bibfnamefont {W.}~\bibnamefont {Malaeb}},
  \bibinfo {author} {\bibfnamefont {Y.}~\bibnamefont {Ishida}}, \bibinfo
  {author} {\bibfnamefont {T.}~\bibnamefont {Sasagawa}}, \bibinfo {author}
  {\bibfnamefont {H.}~\bibnamefont {Sakamoto}}, \bibinfo {author}
  {\bibfnamefont {T.}~\bibnamefont {Takeuchi}}, \bibinfo {author}
  {\bibfnamefont {T.}~\bibnamefont {Tohyama}},\ and\ \bibinfo {author}
  {\bibfnamefont {S.}~\bibnamefont {Shin}},\ }\bibfield  {title} {\bibinfo
  {title} {Point nodes persisting far beyond tc in bi2212},\ }\bibfield
  {journal} {\bibinfo  {journal} {Nature Communications}\ }\textbf {\bibinfo
  {volume} {6}},\ \href {https://doi.org/10.1038/ncomms8699}
  {10.1038/ncomms8699} (\bibinfo {year} {2015})\BibitemShut {NoStop}%
\bibitem [{\citenamefont {Madan}\ \emph {et~al.}(2014)\citenamefont {Madan},
  \citenamefont {Kurosawa}, \citenamefont {Toda}, \citenamefont {Oda},
  \citenamefont {Mertelj}, \citenamefont {Kusar},\ and\ \citenamefont
  {Mihailovic}}]{Madan2014}%
  \BibitemOpen
  \bibfield  {author} {\bibinfo {author} {\bibfnamefont {I.}~\bibnamefont
  {Madan}}, \bibinfo {author} {\bibfnamefont {T.}~\bibnamefont {Kurosawa}},
  \bibinfo {author} {\bibfnamefont {Y.}~\bibnamefont {Toda}}, \bibinfo {author}
  {\bibfnamefont {M.}~\bibnamefont {Oda}}, \bibinfo {author} {\bibfnamefont
  {T.}~\bibnamefont {Mertelj}}, \bibinfo {author} {\bibfnamefont
  {P.}~\bibnamefont {Kusar}},\ and\ \bibinfo {author} {\bibfnamefont
  {D.}~\bibnamefont {Mihailovic}},\ }\bibfield  {title} {\bibinfo {title}
  {Separating pairing from quantum phase coherence dynamics above the
  superconducting transition by femtosecond spectroscopy},\ }\bibfield
  {journal} {\bibinfo  {journal} {Scientific Reports}\ }\textbf {\bibinfo
  {volume} {4}},\ \href {https://doi.org/10.1038/srep05656} {10.1038/srep05656}
  (\bibinfo {year} {2014})\BibitemShut {NoStop}%
\bibitem [{\citenamefont {Reber}\ \emph {et~al.}(2013)\citenamefont {Reber},
  \citenamefont {Plumb}, \citenamefont {Cao}, \citenamefont {Sun},
  \citenamefont {Wang}, \citenamefont {McElroy}, \citenamefont {Iwasawa},
  \citenamefont {Arita}, \citenamefont {Wen}, \citenamefont {Xu}, \citenamefont
  {Gu}, \citenamefont {Yoshida}, \citenamefont {Eisaki}, \citenamefont
  {Aiura},\ and\ \citenamefont {Dessau}}]{Reber2013}%
  \BibitemOpen
  \bibfield  {author} {\bibinfo {author} {\bibfnamefont {T.~J.}\ \bibnamefont
  {Reber}}, \bibinfo {author} {\bibfnamefont {N.~C.}\ \bibnamefont {Plumb}},
  \bibinfo {author} {\bibfnamefont {Y.}~\bibnamefont {Cao}}, \bibinfo {author}
  {\bibfnamefont {Z.}~\bibnamefont {Sun}}, \bibinfo {author} {\bibfnamefont
  {Q.}~\bibnamefont {Wang}}, \bibinfo {author} {\bibfnamefont {K.}~\bibnamefont
  {McElroy}}, \bibinfo {author} {\bibfnamefont {H.}~\bibnamefont {Iwasawa}},
  \bibinfo {author} {\bibfnamefont {M.}~\bibnamefont {Arita}}, \bibinfo
  {author} {\bibfnamefont {J.~S.}\ \bibnamefont {Wen}}, \bibinfo {author}
  {\bibfnamefont {Z.~J.}\ \bibnamefont {Xu}}, \bibinfo {author} {\bibfnamefont
  {G.}~\bibnamefont {Gu}}, \bibinfo {author} {\bibfnamefont {Y.}~\bibnamefont
  {Yoshida}}, \bibinfo {author} {\bibfnamefont {H.}~\bibnamefont {Eisaki}},
  \bibinfo {author} {\bibfnamefont {Y.}~\bibnamefont {Aiura}},\ and\ \bibinfo
  {author} {\bibfnamefont {D.~S.}\ \bibnamefont {Dessau}},\ }\bibfield  {title}
  {\bibinfo {title} {Prepairing and the ``filling'' gap in the cuprates from
  the tomographic density of states},\ }\href
  {https://doi.org/10.1103/PhysRevB.87.060506} {\bibfield  {journal} {\bibinfo
  {journal} {Phys. Rev. B}\ }\textbf {\bibinfo {volume} {87}},\ \bibinfo
  {pages} {060506} (\bibinfo {year} {2013})}\BibitemShut {NoStop}%
\bibitem [{\citenamefont {Yu}\ \emph {et~al.}(2019)\citenamefont {Yu},
  \citenamefont {Xia}, \citenamefont {Pelc}, \citenamefont {He}, \citenamefont
  {Kaneko}, \citenamefont {Sasagawa}, \citenamefont {Li}, \citenamefont {Zhao},
  \citenamefont {Bari\ifmmode \check{s}\else \v{s}\fi{}i\ifmmode~\acute{c}\else
  \'{c}\fi{}}, \citenamefont {Shekhter},\ and\ \citenamefont
  {Greven}}]{Yu2019universal}%
  \BibitemOpen
  \bibfield  {author} {\bibinfo {author} {\bibfnamefont {G.}~\bibnamefont
  {Yu}}, \bibinfo {author} {\bibfnamefont {D.-D.}\ \bibnamefont {Xia}},
  \bibinfo {author} {\bibfnamefont {D.}~\bibnamefont {Pelc}}, \bibinfo {author}
  {\bibfnamefont {R.-H.}\ \bibnamefont {He}}, \bibinfo {author} {\bibfnamefont
  {N.-H.}\ \bibnamefont {Kaneko}}, \bibinfo {author} {\bibfnamefont
  {T.}~\bibnamefont {Sasagawa}}, \bibinfo {author} {\bibfnamefont
  {Y.}~\bibnamefont {Li}}, \bibinfo {author} {\bibfnamefont {X.}~\bibnamefont
  {Zhao}}, \bibinfo {author} {\bibfnamefont {N.}~\bibnamefont {Bari\ifmmode
  \check{s}\else \v{s}\fi{}i\ifmmode~\acute{c}\else \'{c}\fi{}}}, \bibinfo
  {author} {\bibfnamefont {A.}~\bibnamefont {Shekhter}},\ and\ \bibinfo
  {author} {\bibfnamefont {M.}~\bibnamefont {Greven}},\ }\bibfield  {title}
  {\bibinfo {title} {Universal precursor of superconductivity in the
  cuprates},\ }\href {https://doi.org/10.1103/PhysRevB.99.214502} {\bibfield
  {journal} {\bibinfo  {journal} {Phys. Rev. B}\ }\textbf {\bibinfo {volume}
  {99}},\ \bibinfo {pages} {214502} (\bibinfo {year} {2019})}\BibitemShut
  {NoStop}%
\bibitem [{\citenamefont {Pelc}\ \emph {et~al.}(2018)\citenamefont {Pelc},
  \citenamefont {Vučković}, \citenamefont {Grbić}, \citenamefont {Požek},
  \citenamefont {Yu}, \citenamefont {Sasagawa}, \citenamefont {Greven},\ and\
  \citenamefont {Barišić}}]{Pelc2018}%
  \BibitemOpen
  \bibfield  {author} {\bibinfo {author} {\bibfnamefont {D.}~\bibnamefont
  {Pelc}}, \bibinfo {author} {\bibfnamefont {M.}~\bibnamefont {Vučković}},
  \bibinfo {author} {\bibfnamefont {M.~S.}\ \bibnamefont {Grbić}}, \bibinfo
  {author} {\bibfnamefont {M.}~\bibnamefont {Požek}}, \bibinfo {author}
  {\bibfnamefont {G.}~\bibnamefont {Yu}}, \bibinfo {author} {\bibfnamefont
  {T.}~\bibnamefont {Sasagawa}}, \bibinfo {author} {\bibfnamefont
  {M.}~\bibnamefont {Greven}},\ and\ \bibinfo {author} {\bibfnamefont
  {N.}~\bibnamefont {Barišić}},\ }\bibfield  {title} {\bibinfo {title}
  {Emergence of superconductivity in the cuprates via a universal percolation
  process},\ }\bibfield  {journal} {\bibinfo  {journal} {Nature
  Communications}\ }\textbf {\bibinfo {volume} {9}},\ \href
  {https://doi.org/10.1038/s41467-018-06707-y} {10.1038/s41467-018-06707-y}
  (\bibinfo {year} {2018})\BibitemShut {NoStop}%
\bibitem [{\citenamefont {Popčević}\ \emph {et~al.}(2018)\citenamefont
  {Popčević}, \citenamefont {Pelc}, \citenamefont {Tang}, \citenamefont
  {Velebit}, \citenamefont {Anderson}, \citenamefont {Nagarajan}, \citenamefont
  {Yu}, \citenamefont {Požek}, \citenamefont {Barišić},\ and\ \citenamefont
  {Greven}}]{Popevi2018}%
  \BibitemOpen
  \bibfield  {author} {\bibinfo {author} {\bibfnamefont {P.}~\bibnamefont
  {Popčević}}, \bibinfo {author} {\bibfnamefont {D.}~\bibnamefont {Pelc}},
  \bibinfo {author} {\bibfnamefont {Y.}~\bibnamefont {Tang}}, \bibinfo {author}
  {\bibfnamefont {K.}~\bibnamefont {Velebit}}, \bibinfo {author} {\bibfnamefont
  {Z.}~\bibnamefont {Anderson}}, \bibinfo {author} {\bibfnamefont
  {V.}~\bibnamefont {Nagarajan}}, \bibinfo {author} {\bibfnamefont
  {G.}~\bibnamefont {Yu}}, \bibinfo {author} {\bibfnamefont {M.}~\bibnamefont
  {Požek}}, \bibinfo {author} {\bibfnamefont {N.}~\bibnamefont {Barišić}},\
  and\ \bibinfo {author} {\bibfnamefont {M.}~\bibnamefont {Greven}},\
  }\bibfield  {title} {\bibinfo {title} {Percolative nature of the
  direct-current paraconductivity in cuprate superconductors},\ }\bibfield
  {journal} {\bibinfo  {journal} {npj Quantum Materials}\ }\textbf {\bibinfo
  {volume} {3}},\ \href {https://doi.org/10.1038/s41535-018-0115-2}
  {10.1038/s41535-018-0115-2} (\bibinfo {year} {2018})\BibitemShut {NoStop}%
\bibitem [{\citenamefont {Anderson}(1958)}]{Anderson1958}%
  \BibitemOpen
  \bibfield  {author} {\bibinfo {author} {\bibfnamefont {P.~W.}\ \bibnamefont
  {Anderson}},\ }\bibfield  {title} {\bibinfo {title} {Random-phase
  approximation in the theory of superconductivity},\ }\href
  {https://doi.org/10.1103/PhysRev.112.1900} {\bibfield  {journal} {\bibinfo
  {journal} {Phys. Rev.}\ }\textbf {\bibinfo {volume} {112}},\ \bibinfo {pages}
  {1900} (\bibinfo {year} {1958})}\BibitemShut {NoStop}%
\bibitem [{\citenamefont {Ghatak}\ and\ \citenamefont
  {Das}(2018)}]{Ghatak2018}%
  \BibitemOpen
  \bibfield  {author} {\bibinfo {author} {\bibfnamefont {A.}~\bibnamefont
  {Ghatak}}\ and\ \bibinfo {author} {\bibfnamefont {T.}~\bibnamefont {Das}},\
  }\bibfield  {title} {\bibinfo {title} {Theory of superconductivity with
  non-hermitian and parity-time reversal symmetric cooper pairing symmetry},\
  }\href {https://doi.org/10.1103/PhysRevB.97.014512} {\bibfield  {journal}
  {\bibinfo  {journal} {Phys. Rev. B}\ }\textbf {\bibinfo {volume} {97}},\
  \bibinfo {pages} {014512} (\bibinfo {year} {2018})}\BibitemShut {NoStop}%
\bibitem [{\citenamefont {Matsunaga}\ \emph {et~al.}(2017)\citenamefont
  {Matsunaga}, \citenamefont {Tsuji}, \citenamefont {Makise}, \citenamefont
  {Terai}, \citenamefont {Aoki},\ and\ \citenamefont
  {Shimano}}]{Matsunaga2017polarization}%
  \BibitemOpen
  \bibfield  {author} {\bibinfo {author} {\bibfnamefont {R.}~\bibnamefont
  {Matsunaga}}, \bibinfo {author} {\bibfnamefont {N.}~\bibnamefont {Tsuji}},
  \bibinfo {author} {\bibfnamefont {K.}~\bibnamefont {Makise}}, \bibinfo
  {author} {\bibfnamefont {H.}~\bibnamefont {Terai}}, \bibinfo {author}
  {\bibfnamefont {H.}~\bibnamefont {Aoki}},\ and\ \bibinfo {author}
  {\bibfnamefont {R.}~\bibnamefont {Shimano}},\ }\bibfield  {title} {\bibinfo
  {title} {Polarization-resolved terahertz third-harmonic generation in a
  single-crystal superconductor nbn: Dominance of the higgs mode beyond the bcs
  approximation},\ }\href {https://doi.org/10.1103/PhysRevB.96.020505}
  {\bibfield  {journal} {\bibinfo  {journal} {Phys. Rev. B}\ }\textbf {\bibinfo
  {volume} {96}},\ \bibinfo {pages} {020505} (\bibinfo {year}
  {2017})}\BibitemShut {NoStop}%
\bibitem [{\citenamefont {Gabriele}\ \emph {et~al.}(2021)\citenamefont
  {Gabriele}, \citenamefont {Udina},\ and\ \citenamefont
  {Benfatto}}]{Gabriele2021}%
  \BibitemOpen
  \bibfield  {author} {\bibinfo {author} {\bibfnamefont {F.}~\bibnamefont
  {Gabriele}}, \bibinfo {author} {\bibfnamefont {M.}~\bibnamefont {Udina}},\
  and\ \bibinfo {author} {\bibfnamefont {L.}~\bibnamefont {Benfatto}},\
  }\bibfield  {title} {\bibinfo {title} {Non-linear terahertz driving of plasma
  waves in layered cuprates},\ }\bibfield  {journal} {\bibinfo  {journal}
  {Nature Communications}\ }\textbf {\bibinfo {volume} {12}},\ \href
  {https://doi.org/10.1038/s41467-021-21041-6} {10.1038/s41467-021-21041-6}
  (\bibinfo {year} {2021})\BibitemShut {NoStop}%
\bibitem [{\citenamefont {Nadeem}\ \emph {et~al.}(2023)\citenamefont {Nadeem},
  \citenamefont {Fuhrer},\ and\ \citenamefont {Wang}}]{Nadeem2023}%
  \BibitemOpen
  \bibfield  {author} {\bibinfo {author} {\bibfnamefont {M.}~\bibnamefont
  {Nadeem}}, \bibinfo {author} {\bibfnamefont {M.~S.}\ \bibnamefont {Fuhrer}},\
  and\ \bibinfo {author} {\bibfnamefont {X.}~\bibnamefont {Wang}},\ }\bibfield
  {title} {\bibinfo {title} {The superconducting diode effect},\ }\href
  {https://doi.org/10.1038/s42254-023-00632-w} {\bibfield  {journal} {\bibinfo
  {journal} {Nature Reviews Physics}\ }\textbf {\bibinfo {volume} {5}},\
  \bibinfo {pages} {558–577} (\bibinfo {year} {2023})}\BibitemShut {NoStop}%
\bibitem [{\citenamefont {Tsuji}\ and\ \citenamefont
  {Aoki}(2015)}]{Tsuji2015theory}%
  \BibitemOpen
  \bibfield  {author} {\bibinfo {author} {\bibfnamefont {N.}~\bibnamefont
  {Tsuji}}\ and\ \bibinfo {author} {\bibfnamefont {H.}~\bibnamefont {Aoki}},\
  }\bibfield  {title} {\bibinfo {title} {{Theory of Anderson pseudospin
  resonance with Higgs mode in superconductors}},\ }\href
  {https://doi.org/10.1103/PhysRevB.92.064508} {\bibfield  {journal} {\bibinfo
  {journal} {Phys. Rev. B}\ }\textbf {\bibinfo {volume} {92}},\ \bibinfo
  {pages} {064508} (\bibinfo {year} {2015})}\BibitemShut {NoStop}%
\end{thebibliography}
\end{document}